\documentclass[10pt,final, twocolumn]{IEEEtran}
\IEEEoverridecommandlockouts

\usepackage{cite,comment}
\usepackage{mathtools}
\usepackage{hyperref}
\usepackage{graphicx}
\usepackage{enumitem}
\usepackage{tikz}
\usepackage{bm}
\usepackage{amsmath}
\usepackage{algorithm}
\usepackage{algorithmic}
\usepackage{amsmath,amssymb,amsthm}
\usepackage{array}
\usepackage{relsize}
\usepackage{subcaption}
\usepackage{mathtools, nccmath}
\usepackage[labelformat=simple]{subcaption}
\usepackage{derivative}
\usepackage{cancel}
\usepackage[top=.75in, left= 0.625in, right = 0.63in, bottom =1in]{geometry}
\def\BibTeX{{\rm B\kern-.05em{\sc i\kern-.025em b}\kern-.08em
    T\kern-.1667em\lower.7ex\hbox{E}\kern-.125emX}}
		
\newcolumntype{P}[1]{>{\centering\arraybackslash}p{#1}}
\DeclareMathOperator{\C}{\mathbb{C}}
\DeclareMathOperator{\E}{\mathbb{E}}
\DeclareMathOperator{\R}{\mathbb{R}}
\DeclareMathOperator{\N}{\mathcal{N}}

\DeclareMathOperator{\CG}{\mathcal{C}}

\DeclareMathOperator*{\argmin}{arg\,min}
\DeclarePairedDelimiter\abs{\lvert}{\rvert}
\DeclareCaptionSubType * [alph]{table}

\newcommand{\A}{\mathbf{A}}

\newcommand{\h}{\mathbf{H}}
\newcommand{\f}{\mathbf{F}}

\newcommand{\I}{\mathbf{I}}

\newcommand{\fs}{\mathbf{f}}

\newcommand{\as}{\mathbf{a}}

\newcommand{\js}{{\rm j}}
\newcommand{\cn}{\mathsf{c}}

\newcommand{\hs}{\mathbf{h}}

\newcommand{\sn}{\mathsf{s}}

\newcommand{\ysn}{\mathsf{y}}
\newcommand{\ys}{\bm{\mathsf{y}}}
\newcommand{\s}{\bm{\mathsf{s}}}
\newcommand{\ws}{\bm{\mathsf{w}}}
\newcommand{\wsn}{\mathsf{w}}

\newcommand{\hr}{\mathsf{H}}
\newcommand{\tr}{\mathsf{T}}

\newcommand{\bc}{\begin{center}}
\newcommand{\ec}{\end{center}}
\newcommand{\norm}[1]{\left\lVert#1\right\rVert}
\newcommand{\ds}{\displaystyle}
\newcommand{\uprightsubscript}[1]{_{\textnormal{#1}}}
\begingroup\lccode`~=`?\lowercase{\endgroup\let~}\uprightsubscript
\AtBeginDocument{\mathcode`?="8000 }

\newcommand{\tn}[1]{\textnormal{#1}}

\renewcommand{\baselinestretch}{1.0}
\theoremstyle{remark}

\definecolor{dg}{RGB}{255,0,0}

\begin{document}

\title{Physically-Consistent Modeling and Optimization of Non-local RIS-Assisted Multi-User MIMO Communication Systems \\
}
\author{\IEEEauthorblockN{Dilki Wijekoon,~\IEEEmembership{Graduate Student Member,~IEEE}, Amine Mezghani,~\IEEEmembership{Member,~IEEE}, \\George C. Alexandropoulos,~\IEEEmembership{Senior~Member,~IEEE}, and Ekram Hossain,~\IEEEmembership{Fellow,~IEEE}
} 
\thanks{D. Wijekoon, A. Mezghani, and E. Hossain are with the Department of Electrical and Computer Engineering at the University of Manitoba, Canada (emails: wijekood@myumanitoba.ca, \{Amine.Mezghani, Ekram.Hossain\}@umanitoba.ca).} 
\thanks{G.~C.~Alexandropoulos is with the Department of Informatics and Telecommunications, National and Kapodistrian University of Athens, 15784 Athens, Greece (e-mail: alexandg@di.uoa.gr).}
\thanks{Part of the paper was presented at the IEEE International Conference on Communications (ICC) 2024~\cite{icc24}.}
}
\maketitle

\begin{abstract}
Mutual Coupling (MC) emerges as an inherent feature in Reconfigurable Intelligent Surfaces (RISs), particularly, when they are fabricated with sub-wavelength inter-element spacing. Hence, any physically-consistent model of the RIS operation needs to accurately describe MC-induced effects. In addition, the design of the ElectroMagnetic (EM) transmit/receive radiation patterns constitutes another critical factor for efficient RIS operation. The latter two factors lead  naturally to the emergence of non-local RIS structures, whose operation can be effectively described via non-diagonal phase shift matrices. In this paper, we focus on jointly optimizing MC and the radiation patterns in multi-user MIMO communication systems assisted by non-local RISs, which are modeled via the scattering parameters. We particularly present a novel problem formulation for the joint optimization of MC, radiation patterns, and the active and passive beamforming in a physically-consistent manner, considering either reflective or transmissive RIS setups. Differently from the current approaches that design the former two parameters on the fly, we present an offline optimization method which is solved for both considered RIS functionalities. Our extensive simulation results, using both parametric and geometric channel models, showcase the validity of the proposed optimization framework over benchmark schemes, indicating that improved performance is achievable without the need for optimizing MC and the radiation patterns of the RIS on the fly, which can be rather cumbersome.
\end{abstract}

\section{Introduction} 
Reconfigurable Intelligent Surfaces (RISs) constitute a technology that has the potential to revolutionize the performance, coverage, security, and energy efficiency of wireless communication systems \cite{RISsurvey2023,RIS_ISAC_SPM,8811733,9721205,8910627,huang2019reconfigurable,9146875,EURASIP_RIS_all}. They consist of large numbers of elements of ultra-low power consumption, whose responses can be dynamically controlled to change the characteristics of incoming signals. Adjusting the phases of the reflected and refracted (or transmitted) signals, RISs can effectively steer signals in specific directions. This technique, referred to as passive beamforming, enables the creation of multiple concentrated energy beams toward end users \cite{9614196,9610122,9174801,RIS_Illumination}. 

Optimizing the RIS passive beamforming enhances both the strength and richness of signals in communication systems. Furthermore, when passive beamforming is optimized jointly with its active counterpart at the Base tation (BS), it results in even greater improvements in signal quality and strength for intended users, while simultaneously reducing interference caused by unintended users \cite{huang2019reconfigurable,9148961,9474428,10417011,9352967,Alexandropoulos2022Pervasive}. Many studies on RISs have primarily focused on reflective RIS configurations \cite{RIS_DRL,10417011,10096563,9226616,9509394,9531372}. In a reflective RIS configuration, both the BS and users are positioned on the same side of the RIS, known as the reflection space. On the other hand, some research studies have explored transmissive RIS configurations \cite{9855406,10144411,9365009,10242373,10177872,zhang2024design}. In this RIS setups, the BS and users are positioned on opposite sides of the metasurface, enabling it to serve users located in the transmission space.

Choosing an accurate phase shift matrix is critical for RIS configuration. The recent works \cite{9913356,9737373} presented RIS models employing non-diagonal phase shift matrices, also known as non-local RIS structures \cite{10052027}\footnote{We will use the term ``non-local'' instead of ``non-diagonal'' as it is the common term in the original physics literature.}, which outperform conventional RIS models. In \cite{9913356}, the authors coined the term ``beyond diagonal'' phase shift matrix and explained different RIS configurations, where direct physical connections are established among the RIS elements in order to create non-diagonal matrix structures. This study \cite{9913356} investigated RIS models corresponding to fully- and group-connected architectures which result in non-diagonal-full matrix and block diagonal matrix structures, respectively. By considering these RIS configurations, the study in \cite{9913356} revealed that the RIS models with block- and non-diagonal matrix structures outperform those with diagonal matrices (or local RIS) in terms of spectral efficiency. In \cite{9737373}, the authors investigated on reducing the hardware complexity caused by full- and block-diagonal structures, while achieving a better gain as compared to the earlier achieved results for non-diagonal cases. To obtain a corresponding non-diagonal matrix, the typical diagonal matrix was permuted with two row/column permutation matrices. Nonetheless, the proposed ``non-local" RIS architectures are complex to implement due to the requirement of physical connections among the RIS elements. This requires extra hardware resulting in increased wiring complexity and control overhead. 

Non-local RIS structures can also be achieved naturally with minimal overhead using an accurate physically (or electromagnetically)-consistent model, which encounters Mutual Coupling (MC) between pairs of RIS elements and transmit/receive radiation patterns. The ElectroMagnetic (EM) interactions among the RIS elements induces the unavoidable MC effect~\cite{PhysFad,10096563,10417011,8350292,1310320,940505}. When RIS elements are placed in close vicinity to each other, especially with sub-wavelength spacing, MC becomes strong and influences the EM behavior of each element. In addition to element spacing, MC depends on the number of RIS elements and scattering cross-section \cite{rabault2023tacit}. In addition, MC leads to imperfect matching, causing a part of the incoming signal to reflect towards the RIS \cite{10096563,10417011}. This phenomenon, known as multiple reflections, occurs as the wave oscillates back and forth between the RIS and the phase shifters \cite{10096563,10417011}. 

Several works study the impact of MC on RIS-assisted wireless systems \cite{PhysFad,9360851,9525465,9319694,AbrardoAndrea2023AaOo,10096563,10417011}, with the majority of them using the impedance $\mathrm{Z}$-parameters for its characterization~\cite{9360851,9525465,9319694}. These works often use the minimum scattering assumptions to derive the impedance matrices, which is not always a valid assumption. On the other hand, the works~\cite{10096563,AbrardoAndrea2023AaOo} focus on the scattering $\mathrm{S}$-parameters to characterize MC. t is noted that the $\mathrm{Z}$-parameters are related to voltages and currents, while the $\mathrm{S}$-parameters refer to the power ratio of the incident and reflected/transmitted waves \cite{alma99137851470001651}. However, these two parameters have a connection with each other \cite{alma99137851470001651}. Irrespective of the modeling approach, all these works confirm that, when MC is present and taken under consideration in the RIS optimization, the system performance improves. However, none of these works explicitly deals with optimizing MC.

In this paper, we propose a novel approach to simultaneously optimize active and passive beamforming in RIS-assisted multi-user Multiple-Input Multiple-Output (MIMO) systems, while also addressing the optimization of MC and radiation patterns based on $\mathrm{S}$-parameters within a physically-consistent framework. We study separately reflective and transmissive RISs, and design optmal non-local RIS structures. Our design represents a new approach wherein we maintain the simplicity of conventional tunable phase shifters in the dynamic part, while introducing a sophisticated static part based on fundamental laws of physics, resulting in a non-local structure. The proposed method ensures ease of implementation, but also enhances practicality compared to designs relying solely on fully-connected complex tunable phase shifters, which require increased overhead. The contributions of this paper are summarized as follows:
\begin{itemize}
    \item We formulate a novel joint optimization problem for active and passive beamforming, along with the optimization of MC and physically-consistent radiation patterns, for multi-user MIMO downlink transmissions, considering a reflective RIS whose behavior is captured via the scattering $\mathrm{S}$-parameters. The MC and the radiation patterns are designed through an offline optimization approach for a class of channels, rather than optimizing it dynamically on the fly for a given channel realization, as being done up to date with state-of-the-art approaches.
    \item We consider a transmissive RIS model within a physically-consistent setting and optimize its parameters jointly with active beamforming. We particularly assume a fully transmissive mode and optimize the RIS radiation pattern offline for a particular class of channels. Assuming negligible MC, the optimized radiation patterns result in a non-local (non-diagonal) RIS structure.
     \item The proposed offline optimization of MC and radiation patterns introduces a complex nested structure into the formulated problems, involving both outer and inner optimization problems. The inner problem focuses on jointly optimizing the RIS phass profiles and the BS active precoding, while the outer problem involves optimizing MC and radiation patterns based on the scattering $\mathrm{S}$-parameters. The method described in \cite{10096563,10417011} addresses the inner problem for the reflective RIS case, and the approach from \cite{9474428} is applied to solve the inner problem for the transmissive RIS case. A novel solution approach for the outer problems in both setups that is based on the projection gradient descent method is presented.
    \item Our extensive simulation results, considering both parametric and geometric channel models, showcase the robustness of the adopted model and the validity of our RIS designs. It is demonstrated that the performance of multi-user MIMO communication systems assisted with either a non-local reflective RIS or a local transmissive RIS is significantly improved compared to state-of-the-art schemes through the proposed optimization framework integrating engineered MC and radiation patterns. This improvement is achieved even with the proposed offline optimization process for the latter two parameters, emphasizing the effectiveness of the proposed MC-aware optimization approach beyond cumbersome real-time adjustments.
\end{itemize}

The rest of the paper is organized as follows. Section II outlines the system model, assumptions, and optimization problem formulation. Section III describes the proposed solution approach, while Section IV describes the simulation results. Finally, the paper is concluded in Section V.

\textbf{Notations}: Scalar variables are denoted by lowercase letters, while vectors and matrices are respectively represented by small and capital boldface letters (e.g., $\s$ and $\mathbf{S}$). The transpose, conjugate, and conjugate transpose are denoted by $(.)^\tr$, $(.)^*$, and $(.)^\hr$, respectively. Notations $\|.\|?2$ and $\|.\|?F$ stand for the Euclidean and Frobenius norms, respectively, whereas the trace and rank of a matrix $\A$ are represented with $\tn{Tr}(\A)$ and $\tn{Rank}(\A)$, respectively.
The expectation operator is indicated by $\E\lbrace.\rbrace$. The term $\tn{Diag}(\bm a)$ refers to a diagonal matrix whose diagonal elements are the components of $\bm a$. $\otimes$ is the Kronecker product, whereas $\textbf{I}_{M}$ and $\textbf{0}_{M}$ are the $M \times M$ identity and all-zeros matrices, respectively.
\begin{figure}[!t]
\includegraphics[scale=0.24]{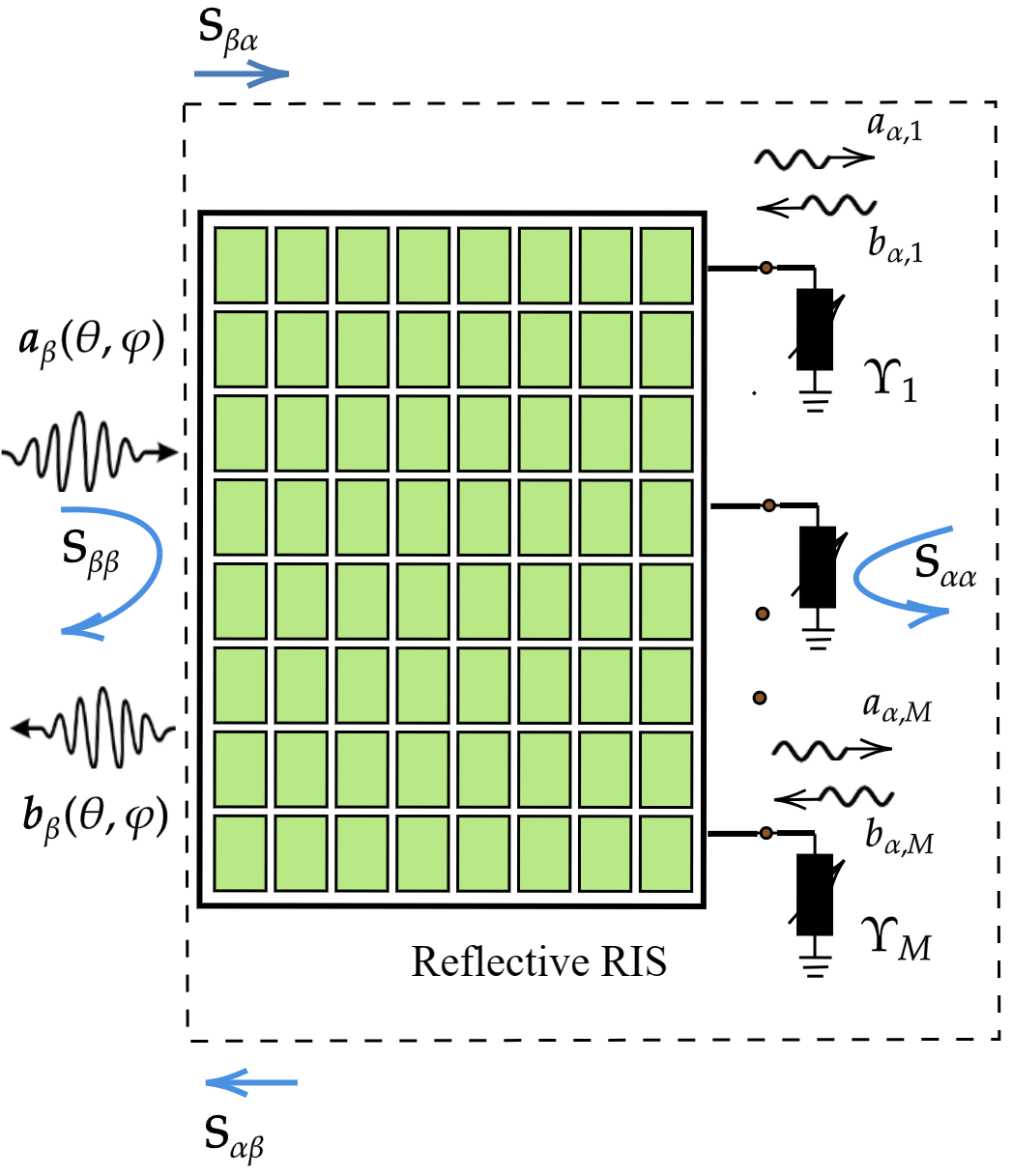}
\caption{Transmit/receive space- and port-side patterns of a reflective RIS.}
\label{fig:1}
\end{figure}
\section{System Model and Problem Formulations}
\subsection{Physically-Consistent Modeling of Reflective RIS}
The EM properties of a RIS are explained by its transmit/receive space- and port-side scattering patterns \cite{10052027}. According to Fig.~\ref{fig:1}, the radiation pattern of the space side is represented by the incoming and outgoing wave phasor vectors $\boldsymbol{a_{\beta}}(\theta,\varphi)$ and $\boldsymbol{b_{\beta}}(\theta,\varphi)$, respectively, with respect to the azimuth and elevation angles $\theta$ and $\varphi$ on the horizontal and vertical axes, respectively. 
Moreover, the forward and backward traveling wave phasors of each element corresponding to the port-side scattering are denoted by $a_{\alpha,m}$ and $b_{\alpha,m}$, respectively, where $m$ (with $m=1,2,\dots,M$) denotes the RIS element index. The mathematical representations of space- and port side-scattering are related as follows: 
\small
\begin{equation}
\label{eqn:sc}
\underbrace{\left[ \begin{array}{c}  \boldsymbol{b_{\beta}}(\theta,\varphi)\\ \boldsymbol{b}_{\alpha} \end{array} \right]}_{\triangleq\mathlarger{\boldsymbol{b}}} = \underbrace{\begin{bmatrix} \mathbf{S_{\beta\beta}} & \mathbf{S}_{\alpha\beta}\\ \mathbf{S}_{\beta\alpha} & \mathbf{S}_{\alpha\alpha} \end{bmatrix}}_{\triangleq\mathlarger{\mathbf{S}}} \underbrace{\left[ \begin{array}{c}  \boldsymbol{a_{\beta}}(\theta,\varphi)\\ \boldsymbol{a}_{\alpha}  \end{array} \right]}_{\triangleq\mathlarger{\boldsymbol{a}}}.
\tag{1}
\end{equation}
\normalsize
In this expression, the matrix $\mathbf{S}$ denotes the total scattering matrix of the RIS. $\mathbf{S}_{\alpha\alpha}$ and $\mathbf{S_{\beta\beta}}$ are called ``multi-port" and ``wave" scattering matrices, respectively. In addition, the matrices $\mathbf{S}_{\beta\alpha}$ and $\mathbf{S}_{\alpha\beta}$ imply transmit and receive radiation patterns indicating MC. The vectors  $\boldsymbol{a_{\alpha}}\triangleq[a_{\alpha,1}, a_{\alpha,2},...,a_{\alpha,M}]^\tr$ and $\boldsymbol{b_{\alpha}}\triangleq[b_{\alpha,1}, b_{\alpha,2},...,b_{\alpha,M}]^\tr$ include, respectively, the incoming and outgoing waves at all RIS port sides. 

In RIS deployments, each port side is terminated by a load. The interaction between incoming and outgoing signals at each port side is characterized by the ``load" scattering matrix $\mathbf{\Upsilon}$, which is represented as follows: 
\small
\begin{equation}
\label{eqn:7-n11}
\boldsymbol{b_{\alpha}}=\mathbf{\Upsilon}\boldsymbol{a_{\alpha}}.
\tag{2}
\end{equation}
\normalsize
Expanding expression (\ref{eqn:sc}) yields the following equations:
\small
\begin{equation}
\label{eqn:7-n1}
\boldsymbol{b_{\beta}}(\theta,\varphi)=\mathbf{S_{\beta\beta}}\boldsymbol{a_{\beta}}(\theta,\varphi)+\mathbf{S}_{\alpha\beta}\boldsymbol{a_{\alpha}},
\tag{3}
\end{equation}
\normalsize
\small
\begin{equation}
\label{eqn:7-n2}
\boldsymbol{b_{\alpha}}=\mathbf{S}_{\beta\alpha}\boldsymbol{a_{\beta}}(\theta,\varphi)+\mathbf{S}_{\alpha\alpha}\boldsymbol{a_{\alpha}}.
\tag{4}
\end{equation}
\normalsize
By utilizing (\ref{eqn:7-n11}), (\ref{eqn:7-n1}), and (\ref{eqn:7-n2}), the total outgoing wave phasor $\boldsymbol{b_{\beta}}(\theta,\varphi)$ at the space-side scattering pattern is given by
\small
\begin{align*}
\label{eqn:e1} 
\boldsymbol{b_{\beta}}(\theta, \varphi)=  \underbrace{\mathbf{S_{\beta \alpha}}\left(\mathbf{\Upsilon}^{-1}-\mathbf{S}_{\alpha\alpha}\right)^{-1}\mathbf{S_{\alpha \beta}}\boldsymbol{a_{\beta}}(\theta,\varphi) }_{\text {Adaptive scattering }}+\underbrace{\mathbf{S_{\beta \beta}}\boldsymbol{a_{\beta}}(\theta,\varphi)}_{\text {Residual scattering}}. \tag {5} 
\end{align*}   
\normalsize
The term $\left(\mathbf{\Upsilon}^{-1}-\mathbf{S}_{\alpha\alpha}\right)^{-1}$ can be expanded as a summation of higher order terms, as shown in the following expression known as ``Neumann series approximation''~\cite{10096563,rabault2023tacit}:
\small
\begin{equation}
\label{eqn:e2}
(\mathbf{\Upsilon}^{-1}-\mathbf{S}_{\alpha\alpha})^{-1}=\ds\sum_{l=0}^\infty(\mathbf{\Upsilon}\mathbf{S}_{\alpha\alpha})^{l}\mathbf{\Upsilon}=\mathbf{\Upsilon}+\mathbf{\Upsilon}\mathbf{S}_{\alpha\alpha}\mathbf{\Upsilon}+\cdots. \tag{6}
\end{equation}
\normalsize
Conventional models utilize a simplified version of (\ref{eqn:e1}) that ignores residual scattering and assume that $\mathbf{S}_{\alpha\beta}=\mathbf{S}_{\beta\alpha}=\mathbf{I}_M$ and absence of multiple reflections, i.e., $\mathbf{S}_{\alpha\alpha}=\mathbf{0}_M$. In this paper, we incorporate MC and multiple reflections \cite{PhysFad,10096563,rabault2023tacit}, and optimize the associated scattering parameters $\mathbf{S}_{\alpha\beta}$, $\mathbf{S}_{\beta\alpha}$, and $\mathbf{S}_{\alpha\alpha}$ for a communication objective. For simplicity, we ignore the residual scattering part in (\ref{eqn:e1}). 
\begin{figure}[!t]
\includegraphics[scale=0.22]{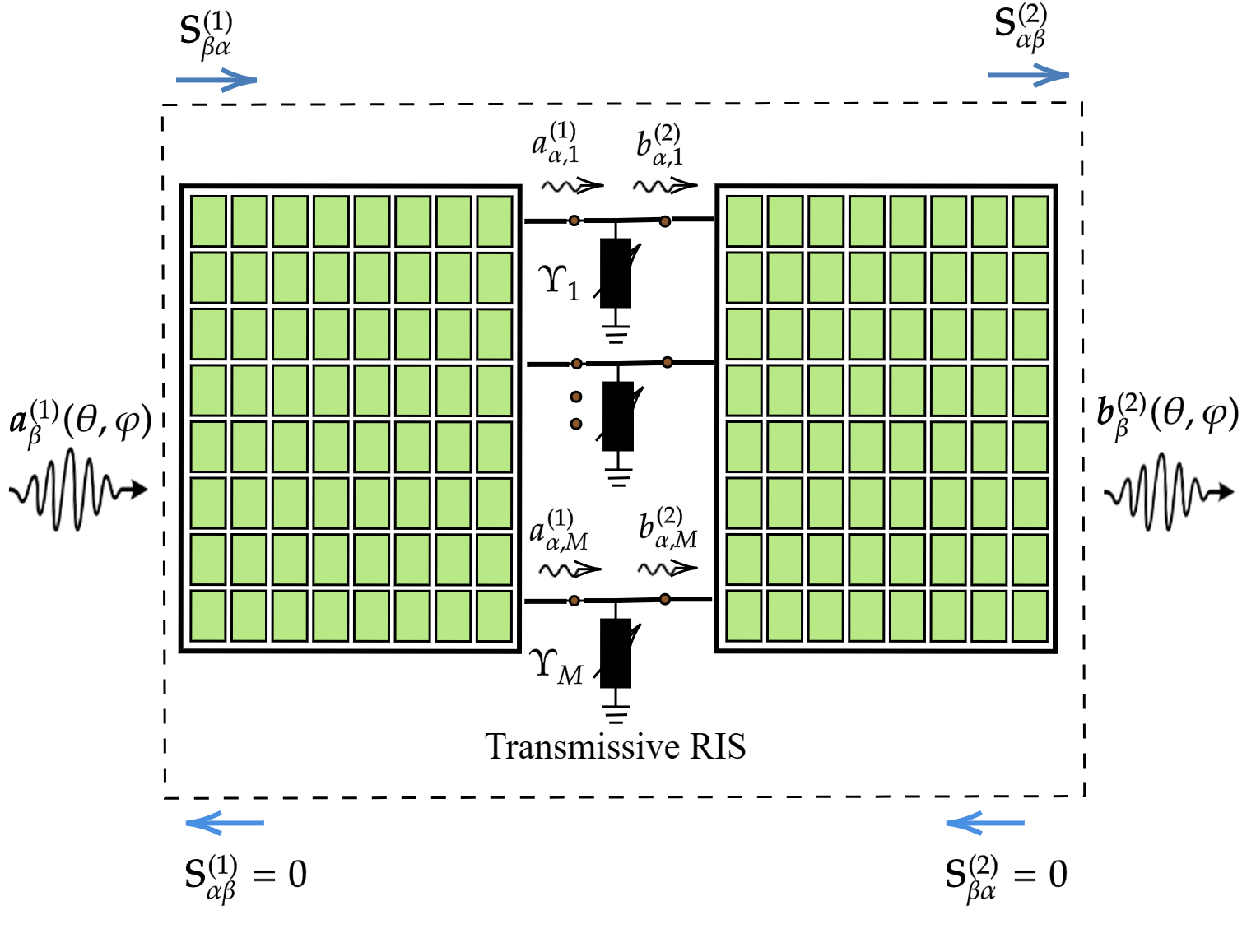}
\caption{Transmit/receive space- and port-side patterns of a transmissive RIS.}
\label{fig:3_trans}
\end{figure}

\subsection{Physically-Consistent Modeling of Transmissive RIS}
Figure \ref{fig:3_trans} displays the space-port side pattern of a transmissive RIS. It contains two arrays connected by phase shifters, with separate space and port sides per array. These port sides are interconnected to enable the functionality of the transmissive RIS. For simplicity, we assume that the RIS is fully transmitting with zero reflection. The space side incoming radiation pattern is denoted by $\boldsymbol{a_{\beta}^{(1)}}(\theta,\varphi)$ while the outgoing radiation pattern is $\boldsymbol{b_{\beta}^{(2)}}(\theta,\varphi)$. 
The forward traveling wave phasor of each element, denoted by $a_{\alpha,m}^{(1)}$, corresponds to the port-side scattering of the first side. Here, the backward traveling wave phasor is zero ($b_{\alpha,m}^{(1)}=0$), indicating no reflection. Conversely, for the second side, the forward traveling wave phasor towards the port side is zero ($a_{\alpha,m}^{(2)}=0$), while the backward traveling wave phasor from the port side $b_{\alpha,m}^{(2)}$ remains. In our ideal transmission scenario, we assume no multiple reflections, resulting in zero multiport scattering matrices for both sides, i.e, $\mathbf{S_{\alpha\alpha}^{(1)}}=\mathbf{S_{\alpha\alpha}^{(2)}}=\textbf{0}_M$. Since the reflection towards the incident side is zero, wave scattering matrices (reflections of space side) are also zero, i.e., $\mathbf{S_{\beta\beta}^{(1)}}=\mathbf{S_{\beta\beta}^{(2)}}=\textbf{0}_M$. In addition, as no signals return to incident directions, matrices $\mathbf{S_{\alpha\beta}^{(1)}}=\mathbf{S_{\beta\alpha}^{(2)}}=\textbf{0}_M$. The equations below represent the space-port sides relationships of the transmissive RIS:
\begin{equation}
\small
\label{eqn:7-n1_7}
\boldsymbol{b_{\beta}^{(2)}}(\theta,\varphi)=\mathbf{S_{\alpha\beta}^{(2)}}\boldsymbol{b_{\alpha}^{(2)}},
\tag{7}
\end{equation}
\normalsize
\small
\begin{equation}
\label{eqn:7-n1_8}
\boldsymbol{a_{\alpha}^{(1)}}=\mathbf{S_{\beta\alpha}^{(1)}}\boldsymbol{a_{\beta}^{(1)}}(\theta,\varphi),
\tag{8}
\end{equation}
\normalsize
\small
\begin{equation}
\label{eqn:7-n11_trans}
\boldsymbol{b_{\alpha}^{(2)}}=\mathbf{\Upsilon}\boldsymbol{a_{\alpha}^{(1)}}.
\tag{9}
\end{equation}
\normalsize
Here the vectors $\boldsymbol{a_{\alpha}^{(1)}}$ and $\boldsymbol{b_{\alpha}^{(2)}}$ are denoted by $\boldsymbol{a_{\alpha}^{(1)}}\triangleq[a_{\alpha,1}^{(1)}, a_{\alpha,2}^{(1)},...,a_{\alpha,M}^{(1)}]^\tr$ and $\boldsymbol{b_{\alpha}^{(2)}}\triangleq[b_{\alpha,1}^{(2)}, b_{\alpha,2}^{(2)},...,b_{\alpha,M}^{(2)}]^\tr$, respectively.
By combining the above equations \eqref{eqn:7-n1_7}, \eqref{eqn:7-n1_8}, and \eqref{eqn:7-n11_trans}, the outgoing wave phasor $\boldsymbol{b_{\beta}^{(2)}}(\theta,\varphi)$ is 
\begin{equation}
\label{eqn:tarns-n1_7}
\boldsymbol{b_{\beta}^{(2)}}(\theta,\varphi)=\mathbf{S_{\alpha\beta}^{(2)}}\mathbf{\Upsilon}\mathbf{S_{\beta\alpha}^{(1)}}\boldsymbol{a_{\beta}^{(1)}}(\theta,\varphi).
\tag{10}
\end{equation}
\normalsize
Hence, in the transmissive RIS scenario, we optimize the scattering parameters $\mathbf{S_{\alpha\beta}^{(2)}}$ and $\mathbf{S_{\beta\alpha}^{(1)}}$, which are traditionally assumed to be identity matrices.

\subsection{Communication System Model}
Consider a multi-user downlink MIMO system comprising a single BS equipped with $N$ antennas, $K<N$ single-antenna users, and a single RIS with $M>K$ elements. The direct channel between the BS and any of the users is assumed to be blocked. The matrix $\h?{r-u} \triangleq [\hs_{\tn{r-u},1}, \hs_{\tn{r-u},2},\ldots, \hs_{\tn{r-u},K}]$ includes all the RIS-user channel gains, with $\hs_{\tn{r-u},k} \in \C^{M}$ representing each RIS-$k$-th user link ($k=1,2,\ldots,K$), while the matrix $\h?{b-r} \in \C^{M\times N}$ corresponds to the BS-RIS channel gains and satisfies the condition: $\tn{Rank}(\h?{b-r})\geq K$.
\begin{figure}
\bc
\scalebox{0.68}{
\begin{picture}(1500,140)
\put(0,0){\includegraphics[scale=0.55]{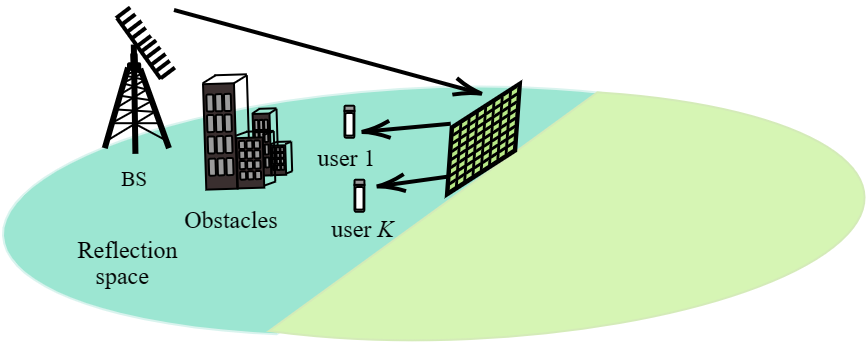}}
\end{picture}
}
\caption{The considered RIS-aided multi-user MIMO system model for reflective RIS.}
\label{fig:2}
\ec
\end{figure}
\vspace{3pt}
\subsubsection{System model for reflective RIS setup}
Figure \ref{fig:2} shows the system model of the reflective RIS setup. It follows from the previous subsection that the effective RIS phase shift matrix is given by $\boldsymbol{\Phi}\triangleq\mathbf{S}_{\beta\alpha}\left(\mathbf{\Upsilon}^{-1}-\mathbf{S}_{\alpha\alpha}\right)^{-1}\mathbf{S}_{\alpha\beta}$, where $\mathbf{\Upsilon}=\tn{Diag}(\bm\upsilon) \in \C^{M \times M}$ with $\bm\upsilon \in \C^M$ including the tunable phase shifts applied to the ports: $\upsilon_m=e^{\js\theta_m}$ ($m=1,2,\ldots,M$) with $\abs*{\upsilon_m}=1$ and $\theta_m\in[0,2\pi]$. Putting all above together, the baseband received signals at all $K$ users can be mathematically expressed as follows:
\begin{equation}
\label{eqn:eq3}
\ys\triangleq\rho\h?{r-u}^\hr\mathbf{S}_{\beta\alpha}{(\mathbf{\Upsilon}^{-1}-\mathbf{S}_{\alpha\alpha})^{-1}}\mathbf{S}_{\alpha\beta}\h?{b-r}\f\s +\rho\ws, \tag {11} 
\end{equation}
where $\ys=[\ysn_1, \ysn_2, \ldots, \ysn_K]^\tr$ and $\s\triangleq[\sn_1,\sn_2,\ldots,\sn_K]^\tr$ represents the transmit symbol vector such that $\sn_k \sim \CG\N(0,1)$ $\forall k$. Here, $\rho \in \R$ is the receiver scaling factor and is assumed to be common for all users. The vector $\ws$ denotes the Additive White Gaussian Noise (AWGN) containing the noise components $\wsn_k \sim \CG\N(0,\sigma?w^2)$. Matrix $\f\triangleq[\fs_1, \fs_2, \ldots, \fs_K]$ with $\fs_k \in \C^{N}$ represents the transmit beamforming matrix for which it holds
$\E_{\s}\big\{\|\f\s\|^2\big\}= P$ with $P$ being the total transmit power. We define the end-to-end RIS-parametrized multi-user channel as follows:
\begin{equation}
\label{eqn:eq5}
\h^\hr\triangleq\h?{r-u}^\hr\mathbf{S}_{\beta\alpha}{(\mathbf{\Upsilon}^{-1}-\mathbf{S}_{\alpha\alpha})^{-1}}\mathbf{S}_{\alpha\beta}\h?{b-r}. \tag {12}
\end{equation}
We now define the Minimum Mean Squared Error (MMSE) criterion \cite{heathmimo,9474428,10096563} which will serve as our system's design optimization objective. The connection between the achievable sum-rate performance and the MMSE of each of the $K$ users can be expressed as follows \cite{9474428,10096563}:
\begin{equation}
\label{eqn:eq4}
C \ = \ \ds\sum_{k=1}^K\log_2\left(\frac{1}{\tn{MMSE}_k}\right) \ = \ \log_2\left(\ds\prod_{k=1}^K\frac{1}{\tn{MMSE}_k}\right), \tag {13} 
\end{equation}
with the total MSE for all $K$ users given by \cite{9474428,10096563}:
\begin{equation}
\label{eqn:eq5_1}
\ds\sum_{k=1}^K\E_{\ysn_k,\sn_k}\left\lbrace\abs*{\ysn_k-\sn_k}^2\right\rbrace \ = \E_{\ys,\s}\left\lbrace\norm{\ys-\s}_2^2\right\rbrace, \tag {14}
\end{equation} 
in which $\E_{\ysn_k,\sn_k}\left\lbrace\abs*{\ysn_k-\sn_k}^2\right\rbrace$ indicates the MSE of the received symbol for each $k$-th user. 
\begin{figure}
\bc
\scalebox{0.68}{
\begin{picture}(1500,140)
\put(18,0){\includegraphics[scale=0.5]{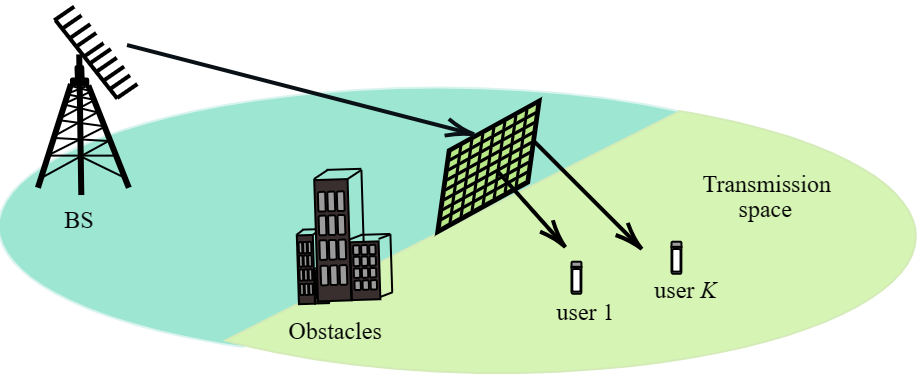}}
\end{picture}
}
\caption{The considered RIS-aided multi-user MIMO system model for transmissive RIS.}
\label{fig:2_trans}
\ec
\end{figure}
\vspace{3pt}
\subsubsection{System model for transmissive RIS setup}
Figure \ref{fig:2_trans} shows the system model of the transmissive RIS setup. According to the previous subsection, the phase shift matrix corresponding to transmissive RIS is given by $\boldsymbol{\Phi}\triangleq\mathbf{S_{\alpha\beta}^{(2)}}\mathbf{\Upsilon}\mathbf{S_{\beta\alpha}^{(1)}}$. Similar to reflective setup, $\mathbf{\Upsilon}=\tn{Diag}(\bm\upsilon) \in \C^{M \times M}$ with $\bm\upsilon \in \C^M$. Even though the matrices $\mathbf{S_{\alpha\alpha}^{(1)}}=\mathbf{S_{\alpha\alpha}^{(2)}}=\textbf{0}_M$, the resulting wave transformation matrix is non-diagonal as it incorporates $\mathbf{S_{\alpha\beta}^{(2)}}$, and $\mathbf{S_{\beta\alpha}^{(1)}}$. Thus, the total received signal can be expressed as:
\begin{equation}
\label{eqn:eq3_trans}
\ys\triangleq\rho\h?{r-u}^\hr\mathbf{S_{\alpha\beta}^{(2)}}\mathbf{\Upsilon}\mathbf{S_{\beta\alpha}^{(1)}}\h?{b-r}\f\s +\ \rho\ws. \tag {15} 
\end{equation}
The vectors $\ys$, $\s$, $\ws$, and the scaling factor $\rho$ carry the same meaning as in the reflective scenario. Furthermore, the matrix $\f$ with $\fs_k \in \C^{N}$ is the transmit beamforming matrix ensuring that $\E\big\{\|\f\s\|^2\big\}= P$. Thereby, the end-to-end RIS-parametrized multi-user channel for transmissive RIS as follows:
\begin{equation}
\label{eqn:eq5_trans}
\h^\hr_{\rm T}\triangleq\h?{r-u}^\hr\mathbf{S_{\alpha\beta}^{(2)}}\mathbf{\Upsilon}\mathbf{S_{\beta\alpha}^{(1)}}\h?{b-r}. \tag {16}
\end{equation}
Note that the MMSE criterion is employed to formulate the problem similarly to the reflective setup.

\subsection{Problem Formulation}
Finding the scattering parameters for each channel that optimize the sum rate in~\eqref{eqn:eq4} on the fly (i.e., in the online phase) is impractical, since this will require changing the RIS accordingly. Therefore, in this paper, we focus on optimizing the scattering matrices $\mathbf{S}_{\alpha\alpha}$, $\mathbf{S}_{\alpha\beta}$, and $\mathbf{S}_{\beta\alpha}$ offline for a particular class of channels corresponding to reflective RIS and the scattering matrices $\mathbf{S_{\alpha\beta}^{(2)}}$ and $\mathbf{S_{\beta\alpha}^{(1)}}$ are optimized offline for a particular class of channels for transmissive RIS. Then, optimize $\f$ and $\mathbf{\Upsilon}$ online (i.e., for each channel realization) for both models. 
\subsubsection{Reflective RIS}
We formulate the following optimization problem for reflective RIS:
\begin{subequations}
\label{eqn:eq6-}
\begin{align}
\label{eqn:eq6-a}
\tag{17a}
\ds\argmin_{ \mathbf{S}_{\alpha\alpha},  \mathbf{S}_{\alpha\beta}, \mathbf{S}_{\beta\alpha}}  \quad & \mathbb{E}_{\h?{r-u},\h?{b-r}}\left[\ds\argmin_{\rho, \f, {\mathbf\Upsilon}} \E_{\ys,\s}\left\lbrace\|\ys-\s\|_2^2\right\rbrace \right], \\ 
\label{eqn:eq6-b}
\tn{subject to} \quad & \tag{17b}
\E_{\s}\left\lbrace\|\f\s\|_2^2\right\rbrace=P,\\ 
\label{eqn:eq6-c} \tag{17c}
& \upsilon_{im}=0, \quad \forall i \neq m,\\ 
\label{eqn:eq6-d} \tag{17d}
& |\upsilon_{ii}|=1, \quad \forall i=1,2, \ldots, M, \\ 
\label{eqn:eq6-e} \tag{17e} 
& \mathbf{S}_{\alpha\alpha}\mathbf{S}_{\alpha\alpha}^\hr+ \mathbf{S}_{\alpha\beta}\mathbf{S}_{\alpha\beta}^\hr=\I_{M}, \\
\label{eqn:eq6-e2} \tag{17f}
& \mathbf{S}_{\alpha\beta}=\mathbf{S}_{\beta\alpha}^\tr \textrm{ and } \mathbf{S}_{\alpha\alpha}=\mathbf{S}_{\alpha\alpha}^\tr.
\end{align}  
\end{subequations}
Constraint \eqref{eqn:eq6-b} is the power allocation constraint, whereas \eqref{eqn:eq6-c} and \eqref{eqn:eq6-d} are unimodular constraints related to the RIS phase shift matrix. Furthermore, the last two constraints stand for the losslessness (conservation of power) and reciprocity of the reflective structure respectively. Using (\ref{eqn:eq3}) and (\ref{eqn:eq5}), the optimization problem is rewritten as:
\begin{subequations}
\label{eqn:eq7}
\begin{align*}
   \tag{18a}
    \label{eqn:eq7-a}
     \ds\argmin_{ \mathbf{S}_{\alpha\alpha},  \mathbf{S}_{\alpha\beta}, \mathbf{S}_{\beta\alpha}} \,\, & \mathbb{E}_{\h?{r-u},\h?{b-r}} \Bigl\{\ds\argmin_{\rho, \f, {\mathbf\Upsilon}}\norm{\rho\h^\hr\f -\I_K}?F^2 + K\rho^2\sigma?w^2\Bigr\}, 
    \vspace{5pt} \\
\label{eqn:eq7-b} \tag{18b}
\tn{subject to} \quad & \|\f\|?F^2=P,\\
\label{eqn:eq7-c} \tag{18c}
& \upsilon_{im}=0, \quad \forall i \neq m,\\
\label{eqn:eq7-d} \tag{18d}
& |\upsilon_{ii}|=1, \quad \forall i=1,2, \ldots, M, \\
\label{eqn:eq7-e} \tag{18e} 
& \mathbf{S}_{\alpha\alpha}\mathbf{S}_{\alpha\alpha}^\hr+ \mathbf{S}_{\alpha\beta}\mathbf{S}_{\alpha\beta}^\hr=\I_{M}, \\
\label{eqn:eq7-e2} \tag{18f}
& \mathbf{S}_{\alpha\beta}=\mathbf{S}_{\beta\alpha}^\tr \textrm{ and } \mathbf{S}_{\alpha\alpha}=\mathbf{S}_{\alpha\alpha}^\tr,
\end{align*}  
\end{subequations}
where $\sigma?w^2$ is the noise variance. 
By incorporating the equality $\mathbf{S}_{\alpha\beta}=\mathbf{S}^\tr_{\beta\alpha}$ into the problem's objective function \eqref{eqn:eq7-a} and using the matrix notation $\hat{\h}^\hr\triangleq\h?{r-u}^\hr\mathbf{S}_{\alpha\beta}^\tr{(\mathbf{\Upsilon}^{-1}-\mathbf{S}_{\alpha\alpha})^{-1}}\mathbf{S}_{\alpha\beta}\h?{b-r}$, yields the following reformulation of the considered optimization problem:
\begin{subequations}
\vspace{-8pt}
\label{eqn:eq8}
\begin{align*}
   \tag{19a}
     \begin{split}
    \label{eqn:eq8-a}
       \ds\argmin_{ \mathbf{S}_{\alpha\alpha},  \mathbf{S}_{\alpha\beta}} \quad & \mathbb{E}_{\h?{r-u},\h?{b-r}} \Bigl\{\ds\argmin_{\rho, \f, {\mathbf\Upsilon}}\norm{\rho\hat{\h}^\hr\f -\I_K}?F^2 + K\rho^2\sigma?w^2 \Bigr\}, 
     \end{split} \\
\label{eqn:eq8-b} \tag{19b}
\tn{subject to} \quad & \|\f\|?F^2=P,\\
\label{eqn:eq8-c} \tag{19c}
& \upsilon_{im}=0, \quad \forall i \neq m,\\
\label{eqn:eq8-d} \tag{19d}
& |\upsilon_{ii}|=1, \quad \forall i=1,2, \cdots, M, \\
\label{eqn:eq8-e} \tag{19e} 
& \mathbf{S}_{\alpha\alpha}\mathbf{S}_{\alpha\alpha}^\hr+ \mathbf{S}_{\alpha\beta}\mathbf{S}_{\alpha\beta}^\hr=\I_{M}, \\
\label{eqn:eq8-f} \tag{19f} 
& \mathbf{S}_{\alpha\alpha}=\mathbf{S}_{\alpha\alpha}^\tr. \end{align*}  
\end{subequations}
Note that, the matrices $\mathbf{S}_{\alpha\alpha}$ and $\mathbf{S}_{\alpha\beta}$ demonstrate partial isometry and exhibit a 2D Toeplitz structure \cite{10052027}. 
Consequently, these matrices can be diagonalized using 2-Discrete Fourier Transform (2-DFT) matrices. This reduces the number of uncoupled ports for the same setup and mitigates beam coupling by avoiding oversampling \cite{10052027}. 
To reduce the complexity of the RIS implementation and for mathematical tractability, we restrict the scattering matrices to be as $\mathbf{S}_{\alpha\alpha}=\textbf{U}\mathbf{\Sigma_{\alpha\alpha}}\textbf{V}^\hr$ and $\mathbf{S}_{\alpha\beta}=\textbf{U}\mathbf{\Sigma_{\alpha\beta}}\textbf{V}^\hr$, where $\mathbf{\Sigma_{\alpha\alpha}}$ and $\mathbf{\Sigma_{\alpha\beta}}$ are complex diagonal matrices whose respective $M$ non-zero elements will be optimized, and the $M\times M$ matrices $\textbf{U}$ and $\textbf{V}$ will be considered as fixed throughout the process. In particular, both $\textbf{U}$ and $\textbf{V}$ are generated as $\mathbf{D}\otimes\mathbf{D}$ with $\mathbf{D}\in\mathbb{C}^{\sqrt{M} \times \sqrt{M}}$ being the DFT matrix, resulting in Toeplitz (i.e., space invariant) RIS structures. The utilization of 2-DFT matrices is also inspired by the concept of cell sectorization, which holds practical significance. Using this simplification and the notation $\Tilde{\mathbf{H}}^\hr\triangleq \h?{r-u}^\hr(\textbf{U}\mathbf{\Sigma}_{\alpha\beta}\textbf{V}^\hr)^\tr{(\mathbf{\Upsilon}^{-1}-\textbf{U}\mathbf{\Sigma}_{\alpha\alpha}\textbf{V}^\hr)^{-1}}(\textbf{U}\mathbf{\Sigma}_{\alpha\beta}\textbf{V}^\hr)\h?{b-r}$, the following formulation is deduced:
\begin{subequations}
\label{eqn:eq9}
\begin{align*}
  \tag{20a}
    \begin{split}
     \label{eqn:eq9-a}
      \ds\argmin_{ \mathbf{\Sigma_{\alpha\alpha}},  \mathbf{\Sigma_{\alpha\beta}}} \quad & \mathbb{E}_{\h?{r-u},\h?{b-r}} \Bigl\{\ds\argmin_{\rho, \f, {\mathbf\Upsilon}}\norm{\rho\Tilde{\mathbf{H}}^\hr\f -\I_K}?F^2 +K\rho^2\sigma?w^2 \Bigr\}, 
    \end{split} \\
    \vspace{5pt}
\label{eqn:eq9-b_1} \tag{20b}
\tn{subject to} \quad & \|\f\|?F^2=P,\\
\label{eqn:eq9-c} \tag{20c}
& \upsilon_{im}=0, \quad \forall i \neq m,\\
\label{eqn:eq9-d} \tag{20d}
& |\upsilon_{ii}|=1, \quad \forall i=1,2, \cdots, M, \\
\label{eqn:eq9-e} \tag{20e} 
&\mathbf{\Sigma_{\alpha\alpha}}\mathbf{\Sigma_{\alpha\alpha}}^\hr+\mathbf{\Sigma_{\alpha\beta}}\mathbf{\Sigma_{\alpha\beta}}^\hr=\I_{M}, \\
\label{eqn:eq9-f} \tag{20f} 
& \textbf{U} \mathbf{\Sigma_{\alpha\alpha}}\textbf{V}^\hr=(\textbf{U} \mathbf{\Sigma_{\alpha\alpha}}\textbf{V}^\hr)^\tr. 
\end{align*}  
\end{subequations}
\subsubsection{Transmissive RIS}
The optimization problem for the transmissive RIS is formulated as follows:
\begin{subequations}
\small
\label{eqn:eq6-_trans}
\begin{align}
\label{eqn:eq6-a_trans}
\tag{21a}
\ds\argmin_{\mathbf{S_{\alpha\beta}^{(2)}}, \mathbf{S_{\beta\alpha}^{(1)}}}  \quad & \mathbb{E}_{\h?{r-u},\h?{b-r}}\left[\ds\argmin_{\rho, \f, {\mathbf\Upsilon}} \E_{\ys,\s}\left\lbrace\|\ys-\s\|_2^2\right\rbrace \right], \\ 
\label{eqn:eq6-b_trans}
\tn{subject to} \quad & \tag{21b}
\E_{\s}\left\lbrace\|\f\s\|_2^2\right\rbrace=P,\\ 
\label{eqn:eq6-c_trans} \tag{21c}
& \upsilon_{im}=0, \quad \forall i \neq m,\\ 
\label{eqn:eq6-d_trans} \tag{21d}
& |\upsilon_{ii}|=1, \quad \forall i=1,2, \ldots, M, \\ 
\label{eqn:eq6-e_trans} \tag{21e} 
& 
\mathbf{S_{\beta\alpha}^{(1)}}\mathbf{S_{\beta\alpha}^{(1)}}^\hr=\I_{M}, \\
\label{eqn:eq6-e2_trans} \tag{21f} 
& 
\mathbf{S_{\alpha\beta}^{(2)}}\mathbf{S_{\alpha\beta}^{(2)}}^\hr=\I_{M}.
\end{align}  
\end{subequations}
By leveraging equations (\ref{eqn:eq3_trans}) and (\ref{eqn:eq5_trans}), the formulated problem for transmissive RIS can be rewritten as follows:
\begin{subequations}
\label{eqn:eq7_trans}
\begin{align*}
   \tag{22a}
    \label{eqn:eq7-a_trans}
     \ds\argmin_{ \mathbf{S_{\alpha\beta}^{(2)}}, \mathbf{S_{\beta\alpha}^{(1)}}} \,\, & \mathbb{E}_{\h?{r-u},\h?{b-r}} \Bigl\{\ds\argmin_{\rho, \f, {\mathbf\Upsilon}}\norm{\rho\h^\hr_{\rm T}\f -\I_K}?F^2 + K\rho^2\sigma?w^2\Bigr\}, 
    \vspace{5pt} \\
\label{eqn:eq7-b_trans} \tag{22b}
\tn{subject to} \quad & \|\f\|?F^2=P,\\
\label{eqn:eq7-c_trans} \tag{22c}
& \upsilon_{im}=0, \quad \forall i \neq m,\\
\label{eqn:eq7-d_trans} \tag{22d}
& |\upsilon_{ii}|=1, \quad \forall i=1,2, \ldots, M, \\
\label{eqn:eq7-e_trans} \tag{22e} 
&  
\mathbf{S_{\beta\alpha}^{(1)}}\mathbf{S_{\beta\alpha}^{(1)}}^\hr=\I_{M}, \\
\label{eqn:eq7-e2_trans} \tag{22f}
& 
\mathbf{S_{\alpha\beta}^{(2)}}\mathbf{S_{\alpha\beta}^{(2)}}^\hr=\I_{M}.
\end{align*}  
\end{subequations}
 The constraints (\ref{eqn:eq7-e_trans}) and (\ref{eqn:eq7-e2_trans}) imply the conservation of power property. Unlike the reflective problem, we are not using any transformations here. We directly proceed with the above-formulated problem.
\section{Proposed EM-Consistent Optimization}
The problems become complex as they involve outer and inner minimization problems and are thus difficult to solve. In this section, we present a decomposition of the overall system optimization into two sub-problems and deploy an alternating optimization approach for both the setups. 

\subsection{Proposed Solution for Reflective RIS}
The two subproblems are given as follows:
\begin{enumerate}
\item For given $\mathbf{\Sigma}_{\alpha\alpha}$ and $\mathbf{\Sigma}_{\alpha\beta}$, solve:
\begin{subequations}
\label{eqn:ao-precod-mat_1}
\begin{align*}
\label{eqn:ao-precod-mata} \tag{23a}
\argmin_{\rho, \f,{\mathbf\Upsilon}} \quad & \norm{\rho\Tilde{\mathbf{H}}^\hr\f -\I_K}?F^2 +K\rho^2\sigma?w^2 \\\
\label{eqn:ao-precod-matb}\tag{23b}
\tn{subject to} \quad & \|\f\|?F^2=P,\\ 
\label{eqn:ao_2} \tag{23c}
& \upsilon_{im}=0, \quad \forall i \neq m,\\
\label{eqn:ao_2_d} \tag{23d}
& |\upsilon_{ii}|=1, \quad \forall i=1,2, \ldots, M. \\
\end{align*}
\end{subequations}
\item For given $\rho$, $\f$, and ${\mathbf\Upsilon}$, solve:
\begin{subequations}
\label{eqn:ao-phase-mat_1}
\begin{align*}
\label{eqn:ao-phase-mata} \tag{24a}
\ds\argmin_{ \mathbf{\Sigma}_{\alpha\alpha},  \mathbf{\Sigma}_{\alpha\beta}} \quad & \mathbb{E}_{\h?{r-u},\h?{b-r}} \Bigl\{\norm{\rho\Tilde{\mathbf{H}}^\hr\f -\I_K}?F^2 +K\rho^2\sigma?w^2\Bigr\} \\\
\label{eqn:eq91-b}\tag{24b}
\tn{subject to} \quad & \mathbf{\Sigma}_{\alpha\alpha}\mathbf{\Sigma}_{\alpha\alpha}^\hr+\mathbf{\Sigma}_{\alpha\beta}\mathbf{\Sigma}_{\alpha\beta}^\hr=\I_{M}, \\
\label{eqn:eq91-c} \tag{24c}
& \textbf{U} \mathbf{\Sigma_{\alpha\alpha}}\textbf{V}^\hr=(\textbf{U} \mathbf{\Sigma_{\alpha\alpha}}\textbf{V}^\hr)^\tr.
\end{align*}
\end{subequations}
\end{enumerate}
The first sub-problem can be efficiently solved via the method described in \cite{10096563,10417011} based on the gradient descent algorithm. The solution to the second sub-problem, which gives the optimal scattering matrices, is next obtained via a gradient step and a closed-form projection step for both $\mathbf{\Sigma_{\alpha\alpha}}$ and $\mathbf{\Sigma_{\alpha\beta}}$ matrices. 
Recall from the previous section that we have used the assumptions $\mathbf{S}_{\alpha\alpha}=\textbf{U}\mathbf{\Sigma_{\alpha\alpha}}\textbf{V}^\hr$ and $\mathbf{S}_{\alpha\beta}=\textbf{U}\mathbf{\Sigma_{\alpha\beta}}\textbf{V}^\hr$, where $\mathbf{\Sigma_{\alpha\alpha}}$ and $\mathbf{\Sigma_{\alpha\beta}}$ are complex diagonal matrices and $\textbf{U}$ and $\textbf{V}$ are fixed DFT-based matrices. Hence, the optimization of the second sub-problem is conducted with respect to $\mathbf{\Sigma_{\alpha\alpha}}$ and $\mathbf{\Sigma_{\alpha\beta}}$ instead of their corresponding scattering parameters. 

\subsubsection{Projected Gradient Descent}
For each channel realization, we define the following MSE function:
\begin{equation}
\textit{f}(\mathbf{\Sigma_{\alpha\alpha}},\mathbf{\Sigma_{\alpha\beta}})\triangleq  \norm{\rho\Tilde{\mathbf{H}}^\hr\f -\I_K}?F^2 +K\rho^2\sigma?w^2. \tag{25}
\end{equation}
Since the objective function (\ref{eqn:ao-phase-mata}) involves an expectation with respect to $\h?{r-u}$ and $\h?{b-r}$, Monte Carlo sampling can be used to compute the gradients for a set of $Q$ channel samples. To this end, the overall gradient of this objective function with respect to $\mathbf{\Sigma_{\alpha\alpha}}$ is computed as follows:
\begin{equation}\label{eq:G_aa}
\mathbf{G}_{\mathbf{\Sigma_{\alpha\alpha}}}\triangleq\frac{1}{Q}\sum_{q=1}^Q \left[\pdv{\textit{f}(\mathbf{\Sigma_{\alpha\alpha}},\mathbf{\Sigma_{\alpha\beta}})}{\mathbf{\Sigma_{\alpha\alpha}}}\right]_\textit{q},\tag{26}
\end{equation}
where the gradient for each $q$-th channel sample is given by:
\begin{align}
\scriptsize
\begin{split}
&\left[\pdv{\textit{f}(\mathbf{\Sigma_{\alpha\alpha}},\mathbf{\Sigma_{\alpha\beta}})}{\mathbf{\Sigma_{\alpha\alpha}}}\right]_\textit{q}=2{ \textbf{U}^\tr}(\mathbf{\Upsilon}_\textit{q}^{-1}-\textbf{U}\mathbf{\Sigma_{\alpha\alpha}}\textbf{V}^\hr)^{-\tr} \left(\rho_\textit{q}\h?{r-u}^\hr(\textbf{U}\mathbf{\Sigma_{\alpha\beta}}\textbf{V}^\hr)^\tr \right)^\tr \\
& \left(\rho_\textit{q}\h?{r-u}^\hr(\textbf{U}\mathbf{\Sigma_{\alpha\beta}}\textbf{V}^\hr)^\tr{(\mathbf{\Upsilon}_\textit{q}^{-1}-\textbf{U}\mathbf{\Sigma_{\alpha\alpha}}\textbf{V}^\hr)^{-1}}(\textbf{U}\mathbf{\Sigma_{\alpha\beta}}\textbf{V}^\hr)\h?{b-r}\f_\textit{q}-\I_K\right)^* \\
& \left((\textbf{U}\mathbf{\Sigma_{\alpha\beta}}\textbf{V}^\hr)\h?{b-r}\f_\textit{q}\right)^{\tr}(\mathbf{\Upsilon}_{\textit{q}}^{-1}-\textbf{U}\mathbf{\Sigma_{\alpha\alpha}}\textbf{V}^\hr)^{-\tr} {\textbf{V}^*},
\end{split}
\tag{27}
\end{align}
with $\mathbf{\Upsilon}_\textit{q}$, $\f_\textit{q}$, and $\rho_\textit{q}$ indicating the optimum RIS phase configuration, BS precoding matrix, and scaling parameter values for each $q$-th channel sample. Similarly, the gradient of (\ref{eqn:ao-phase-mata}) with respect to $\mathbf{\Sigma_{\alpha\beta}}$ can be approximated as: 
 \begin{equation}\label{eq:G_ab}  
 \mathbf{G}_{\mathbf{\Sigma_{\alpha\beta}}}\triangleq
\frac{1}{Q}\sum_{q=1}^Q \left[\pdv{\textit{f}(\mathbf{\Sigma_{\alpha\alpha}},\mathbf{\Sigma_{\alpha\beta}})}{\mathbf{\Sigma_{\alpha\beta}}}\right]_\textit{q}, \tag{28}
\end{equation}
where the gradient for each $q$-th channel sample is given by:
\begin{align}
\scriptsize
\begin{split}
& \left[\pdv{\textit{f}(\mathbf{\Sigma_{\alpha\alpha}},\mathbf{\Sigma_{\alpha\beta}})}{\mathbf{\Sigma_{\alpha\beta}}}\right]_\textit{q}=2 \left(\rho_\textit{q}\h?{r-u}^\hr(\textbf{V}^\hr)^\tr \right)^\tr \\
& \left(\rho_\textit{q}\h?{r-u}^\hr(\textbf{U}\mathbf{\Sigma_{\alpha\beta}}\textbf{V}^\hr)^\tr{(\mathbf{\Upsilon}_{\textit{q}}^{-1}-\textbf{U}\mathbf{\Sigma_{\alpha\alpha}}\textbf{V}^\hr)^{-1}}(\textbf{U}\mathbf{\Sigma_{\alpha\beta}}\textbf{V}^\hr)\h?{b-r}\f_{\textit{q}}-\I_K\right)^* \\
& (\textbf{V}^\hr\h?{b-r}\f_{\textit{q}})^{\tr}\mathbf{\Sigma_{\alpha\beta}}^\tr
\left(\textbf{U}^\tr(\mathbf{\Upsilon}_{\textit{q}}^{-1}-\textbf{U}\mathbf{\Sigma_{\alpha\alpha}}\textbf{V}^\hr)^{-1}\textbf{U}\right)^\tr  + \\
& 2\left(\textbf{U}^\tr(\mathbf{\Upsilon}_{\textit{q}}^{-1}-\textbf{U}\mathbf{\Sigma_{\alpha\alpha}}\textbf{V}^\hr)^{-1}\textbf{U}\right)^\tr\mathbf{\Sigma_{\alpha\beta}}^\tr\left(\rho_\textit{q}\h?{r-u}^\hr(\textbf{V}^\hr)^\tr\right)^\tr \\ & \left(\rho_\textit{q}\h?{r-u}^\hr(\textbf{U}\mathbf{\Sigma_{\alpha\beta}}\textbf{V}^\hr)^\tr{(\mathbf{\Upsilon}_{\textit{q}}^{-1}-\textbf{U}\mathbf{\Sigma_{\alpha\alpha}}\textbf{V}^\hr)^{-1}}(\textbf{U}\mathbf{\Sigma_{\alpha\beta}}\textbf{V}^\hr)\h?{b-r}\f_{\textit{q}}-\I_K\right)^* \\
& (\textbf{V}^\hr\h?{b-r}\f_{\textit{q}})^{\tr}.
\end{split}
\tag{29}
\end{align}
Once the gradients are computed, the parameters $\mathbf{\Sigma_{\alpha\alpha}}$ and $\mathbf{\Sigma_{\alpha\beta}}$ are updated as follows:
\begin{align}
\mathbf{\Tilde{\Sigma}_{\alpha\alpha}}&=\mathbf{\Sigma_{\alpha\alpha}}-\mu{\bf Diag}\left\{\mathbf{G}^{*}_{\mathbf{\Sigma_{\alpha\alpha}}}\right\}, \tag{30}
\\
\mathbf{\Tilde{\Sigma}_{\alpha\beta}}&=\mathbf{\Sigma_{\alpha\beta}}-\mu{\bf  Diag}\left\{\mathbf{G}^{ *}_{\mathbf{\Sigma_{\alpha\beta}}}\right\}, \tag{31}
\end{align}
where we have adopted the conjugate of the gradients as the search directions and $\mu$ denotes the fixed step size. To maintain the symmetry of $\mathbf{\Tilde{\Sigma}_{\alpha\alpha}}$ as per (\ref{eqn:eq91-c}), we symmetrically update its diagonal elements at positions $\Tilde{\sigma}_{\alpha\alpha,ii}$ and $\Tilde{\sigma}_{\alpha\alpha,(M-i+{2})(M-i+{2})}$, $\forall$$i={2,3,\ldots,M/2}$ by taking their average, as follows:
\begin{align}
\label{eq:symmetric}
&\Tilde{\sigma}_{\alpha\alpha,ii}=\frac{\Tilde{\sigma}_{\alpha\alpha,ii}+ \Tilde{\sigma}_{\alpha\alpha,(M-i+{2})(M-i+{2})}}{2} \tag{32},\\
\label{eq:symmetric_1}
&\Tilde{\sigma}_{\alpha\alpha,(M-i+{2})(M-i+{2})}=\frac{\Tilde{\sigma}_{\alpha\alpha,ii}+\Tilde{\sigma}_{\alpha\alpha,(M-i+{2})(M-i+{2})}}{2} \tag{33},
\end{align}
Clearly, when $\mathbf{\Sigma_{\alpha\alpha}}$ is symmetric, $\mathbf{\Sigma_{\alpha\beta}}$ becomes symmetric as well since $\mathbf{I}_M$ is also symmetric. Therefore, we also make $\mathbf{\Tilde{\Sigma}_{\alpha\beta}}$ symmetric using similar expressions to (\ref{eq:symmetric}) and (\ref{eq:symmetric_1}). 

We now formulate the following projection problem that considers both constraints \eqref{eqn:eq91-b} and \eqref{eqn:eq91-c} for the design of symmetric $\mathbf{\Sigma_{\alpha\alpha}}$ and $\mathbf{\Sigma_{\alpha\beta}}$:
\begin{subequations}
\label{eqn:pgg} 
\begin{align*}
   \tag{34a}
    \begin{split}
     \label{eqn:pg}
     \ds\argmin_{ \mathbf{\Sigma_{\alpha\alpha}},  \mathbf{\Sigma_{\alpha\beta}}} \quad & \norm{\mathbf{{\Sigma}}_{\alpha\alpha}-\mathbf{\Tilde{\Sigma}}_{\alpha\alpha}}?F^2 +\norm{\mathbf{\Sigma}_{\alpha\beta}-\mathbf{\Tilde{\Sigma}}_{\alpha\beta}}?F^2
    \end{split} \\
    \vspace{5pt}
\label{eqn:pg_1} \tag{34b} 
\tn{subject to} \quad &
\mathbf{\Sigma}_{\alpha\alpha}\mathbf{\Sigma}_{\alpha\alpha}^\hr+\mathbf{\Sigma}_{\alpha\beta}\mathbf{\Sigma}_{\alpha\beta}^\hr=\I_{M}. \end{align*}  
\end{subequations} 
We then simplify the term $\norm{\mathbf{\Sigma_{\alpha\alpha}}-\mathbf{\Tilde{\Sigma}_{\alpha\alpha}}}?F^2$ as follows:
\small
\begin{align} \tag{35}
\label{eqn:eq333}
\begin{split}
        &\norm{\mathbf{\Sigma_{\alpha\alpha}}-\mathbf{\Tilde{\Sigma}_{\alpha\alpha}}}?F^2 ~=~ \tn{Tr}\bigl((\mathbf{\Sigma_{\alpha\alpha}}-\mathbf{\Tilde{\Sigma}_{\alpha\alpha}})(\mathbf{\Sigma_{\alpha\alpha}}-\mathbf{\Tilde{\Sigma}_{\alpha\alpha}})^\hr\bigr)\\ &
        ~=~ \sum_{i=1}^{M} {|}\sigma_{\alpha\alpha,ii}{|}^2 -2{\rm Re\{}\sigma_{\alpha\alpha,ii}\Tilde{\sigma}_{\alpha\alpha,ii}{^*\}}+{|}\Tilde{\sigma}_{\alpha\alpha,ii}{|}^2,
    \end{split}
\end{align}
\normalsize
where $\sigma_{\alpha\alpha,ii}$ and $\Tilde{\sigma}_{\alpha\alpha,ii}$ are the diagonal elements of $\mathbf{\Sigma_{\alpha\alpha}}$ and $\mathbf{\Tilde{\Sigma}_{\alpha\alpha}}$, respectively. Similarly, the other term $\norm{\mathbf{\Sigma_{\alpha\beta}}-\mathbf{\Tilde{\Sigma}_{\alpha\beta}}}?F^2$ in the objective can be simplified as:
\small
\begin{align} \tag{36}
\label{eqn:eq333_1}
\begin{split}
        &\norm{\mathbf{\Sigma_{\alpha\beta}}-\mathbf{\Tilde{\Sigma}_{\alpha\beta}}}?F^2 ~=~ \tn{Tr}\bigl((\mathbf{\Sigma_{\alpha\beta}}-\mathbf{\Tilde{\Sigma}_{\alpha\beta}})(\mathbf{\Sigma_{\alpha\beta}}-\mathbf{\Tilde{\Sigma}_{\alpha\beta}})^\hr\bigr)\\ &
        ~=~ \sum_{i=1}^{M} {|}\sigma_{\alpha\beta,ii}{|}^2 -2{\rm Re\{}\sigma_{\alpha\beta,ii}\Tilde{\sigma}_{\alpha\beta,ii}{^*\}}+{|}\Tilde{\sigma}_{\alpha\beta,ii}{|}^2,
    \end{split}
\end{align}
\normalsize
where the scalars $\sigma_{\alpha\beta,ii}$ and $\Tilde{\sigma}_{\alpha\beta,ii}$ represent the diagonal elements of the matrices $\mathbf{\Sigma_{\alpha\beta}}$ and $\mathbf{\Tilde{\Sigma}_{\alpha\beta}}$, respectively. Using the latter expressions, the projection problem in \eqref{eqn:pg} and \eqref{eqn:pg_1} can be reformulated as follows:
\begin{subequations}
\label{eqn:eq_new_1}
\begin{align}
   \tag{37a}
    \begin{split}
     \label{eqn:new_1}
      \argmin_{ \sigma_{\alpha\alpha,ii},  \sigma_{\alpha\beta,ii}} \,\, &\sum_{i=1}^{M} ({|}\sigma_{\alpha\alpha,ii}{|}^2 -2{\rm Re\{}\sigma_{\alpha\alpha,ii}\Tilde{\sigma}_{\alpha\alpha,ii}{^*\}}+{|}\Tilde{\sigma}_{\alpha\alpha,ii}{|}^2) \\ &+ \sum_{i=1}^{M} ({|}\sigma_{\alpha\beta,ii}{|}^2 -2{\rm Re\{}\sigma_{\alpha\beta,ii}\Tilde{\sigma}_{\alpha\beta,ii}{^*\}}+{|}\Tilde{\sigma}_{\alpha\beta,ii}{|}^2)
    \end{split} \\
\label{eqn:new_2} \tag{37b} 
\tn{subject to} \,\,
& {|}\sigma_{\alpha\alpha,ii}{|}^2+{|}\sigma_{\alpha\beta,ii}{|}^2=1,  \,\, \forall i=1,2,\ldots,M.
\end{align}  
\end{subequations} 

In the sequel, we will present closed-form solutions for the latter problem using the method of Lagrange multipliers.

\subsubsection{Closed-Form Solution}
The Lagrangian function of the latter optimization problem is defined as follows:
\begin{align} 
 &\nonumber L(\sigma_{\alpha\alpha,ii},\sigma_{\alpha\beta,ii},\lambda_i)\triangleq \\ & \sum_{i=1}^{M} ({|}\sigma_{\alpha\alpha,ii}{|}^2 -2{\rm Re\{}\sigma_{\alpha\alpha,ii}\Tilde{\sigma}_{\alpha\alpha,ii}{^*\}}+{|}\Tilde{\sigma}_{\alpha\alpha,ii}{|}^2) \nonumber \\ &  +\sum_{i=1}^{M} ({|}\sigma_{\alpha\beta,ii}{|}^2 -2{\rm Re\{}\sigma_{\alpha\beta,ii}\Tilde{\sigma}_{\alpha\beta,ii}{^*\}}+{|}\Tilde{\sigma}_{\alpha\beta,ii}{|}^2)\nonumber  \\ & +\sum_{i=1}^{M}\lambda_i({|}\sigma_{\alpha\alpha,ii}{|}^2 +{|}\sigma_{\alpha\beta,ii}{|}^2 -1).\tag{38}\nonumber 
\end{align}
\normalsize
Its partial derivatives with respect to the optimization variables $\sigma_{\alpha\alpha,ii}$, $\sigma_{\alpha\beta,ii}$, and $\lambda_i$ are then computed as:
\begin{align} \tag{39}
\label{eqn:eq333_1_1}
&\pdv{L(\sigma_{\alpha\alpha,ii},\sigma_{\alpha\beta,ii},\lambda_i)}{\sigma_{\alpha\alpha,ii}}=
2\sigma_{\alpha\alpha,ii}-2\Tilde{\sigma}_{\alpha\alpha,ii}+2\lambda_i\sigma_{\alpha\alpha,ii},
\end{align}
\normalsize
\begin{align} \tag{40}
\label{eqn:eq333_1_2}
&\pdv{L(\sigma_{\alpha\alpha,ii},\sigma_{\alpha\beta,ii},\lambda_i)}{\sigma_{\alpha\beta,ii}}= 2\sigma_{\alpha\beta,ii}-2\Tilde{\sigma}_{\alpha\beta,ii}+2\lambda_i\sigma_{\alpha\beta,ii},
\end{align}
\normalsize
\begin{align} \tag{41}
\label{eqn:eq333_1_3}
&\pdv{L(\sigma_{\alpha\alpha,ii},\sigma_{\alpha\beta,ii},\lambda_i)}{\lambda_i} = {|}\sigma_{\alpha\alpha,ii}{|}^2+{|}\sigma_{\alpha\beta,ii}{|}^2-1. 
\end{align}
\normalsize
By equating each of the latter expressions to zero and then solving the resulting $3\times3$ system of equations, the diagonal elements of $\mathbf{\Sigma_{\alpha\alpha}}$ and $\mathbf{\Sigma_{\alpha\beta}}$ are obtained in closed-form as:
\begin{equation} \tag{42}
\label{eq:sol1_1}
\sigma_{\alpha\alpha,ii}=\frac{\Tilde{\sigma}_{\alpha\alpha,ii}}{\sqrt{{|}\Tilde{\sigma}_{\alpha\alpha,ii}{|}^2+{|}\Tilde{\sigma}_{\alpha\beta,ii}{|}^2}},
\end{equation}
\begin{equation} \tag{43}
\label{eq:sol1}
\sigma_{\alpha\beta,ii}=\frac{\Tilde{\sigma}_{\alpha\beta,ii}}{\sqrt{ {|}\Tilde{\sigma}_{\alpha\alpha,ii}{|}^2+{|}\Tilde{\sigma}_{\alpha\beta,ii}{|}^2}}.
\end{equation}
The solution of the overall system optimization problem is summarized in \textbf{Algorithm} \ref{algo:joint_op}.
\begin{algorithm}[!t]
\caption{System Optimization for Reflective RIS}
\label{algo:joint_op}
\begin{algorithmic}[1]
\renewcommand{\algorithmicrequire}{\textbf{Input:}}
\renewcommand{\algorithmicensure}{\textbf{Output:}}
\REQUIRE $Q$ samples of the channel matrices $\h?{b-r}$ and $\h?{r-u}$.
\STATE Set $\textbf{U}=\textbf{V}=\mathbf{D} \otimes\mathbf{D}$.
\vspace{1pt}
\STATE Initialize $\widehat{\mathbf{\Sigma}}_{\alpha\alpha_{0}}$ and $\widehat{\mathbf{\Sigma}}_{\alpha\beta_{0}}$. 
\FOR{$k=1,2,\ldots,I_{\max}$}
\FOR{$q=1,2,\ldots,Q$}
\STATE Initialize $\mathbf{\Upsilon}_k$.
\STATE Initialize $\rho_k$ and $\f_k$ as described in \cite{10096563}.
\STATE Compute $\mathbf{\Upsilon}_k$ and $\f_k$ solving sub-problem 1) via the method in \cite{10096563}.
\STATE Compute $\left[\pdv{\textit{f}(\mathbf{\Sigma}_{\alpha\alpha,k},\mathbf{\Sigma}_{\alpha\beta,k})}{\mathbf{\Sigma}_{\alpha\alpha,k}}\right]_\textit{q}$.
\STATE Compute $\left[\pdv{\textit{f}(\mathbf{\Sigma}_{\alpha\alpha,k},\mathbf{\Sigma}_{\alpha\beta,k})}{\mathbf{\Sigma}_{\alpha\beta,k}}\right]_{\textit{q}}$.
\ENDFOR
\STATE Compute $\mathbf{G}_{\mathbf{\Sigma}_{\alpha\alpha,k}}$ using \eqref{eq:G_aa}.  
\STATE Compute $\mathbf{G}_{\mathbf{\Sigma}_{\alpha\beta,k}}$ using \eqref{eq:G_ab}.
\STATE Compute $\mathbf{\Sigma}_{\alpha\alpha,k}=\mathbf{\Sigma}_{\alpha\alpha,k-1}-\mu{\bf Diag}\left\{\mathbf{G}^{*}_{\mathbf{\Sigma}_{\alpha\alpha,k}}\right\}$.
\STATE Compute 
$\mathbf{\Sigma}_{\alpha\beta,k}=\mathbf{\Sigma}_{\alpha\beta,k-1}-\mu{\bf Diag}\left\{\mathbf{G}^{*}_{\mathbf{\Sigma}_{\alpha\beta,k}}\right\}$.
\vspace{1.5pt}
\STATE Use (\ref{eq:symmetric}) and (\ref{eq:symmetric_1}) to symmetrify $\mathbf{\Sigma}_{\alpha\alpha,k}$ and $\mathbf{\Sigma}_{\alpha\beta,k}$.
\vspace{1.5pt}
\STATE Compute $\widehat{\sigma}_{\alpha\alpha,ii,k}=\frac{\sigma_{\alpha\alpha,ii,k}}{\sqrt{|\sigma_{\alpha\alpha,ii,k}|^2+|\sigma_{\alpha\beta,ii,k}|^2}}$.
\vspace{1.5pt}
\STATE Compute $\widehat{\sigma}_{\alpha\beta,ii,k}=\frac{\sigma_{\alpha\beta,ii,k}}{\sqrt{|\sigma_{\alpha\alpha,ii,k}|^2+|\sigma_{\alpha\beta,ii,k}|^2}}$.
\vspace{1.5pt}
\STATE Set $\mathbf{S}_{\alpha\alpha,k}=\textbf{U}\tn{Diag}\left(\left[\widehat{\sigma}_{\alpha\alpha,11,k},\ldots,\widehat{\sigma}_{\alpha\alpha,MM,k}\right]\right)\textbf{V}^\hr$.
\STATE Set $\mathbf{S}_{\alpha\beta,k}=\textbf{U}\tn{Diag}\left(\left[\widehat{\sigma}_{\alpha\beta,11,k},\ldots,\widehat{\sigma}_{\alpha\beta,MM,k}\right]\right)\textbf{V}^\hr$.
\ENDFOR
\ENSURE $\rho_{I_{\max}}$, $\f_{I_{\max}}$, $\mathbf{\Upsilon}_{I_{\max}}$, $\mathbf{S}_{\alpha\alpha,I_{\max}}$, and $\mathbf{S}_{\alpha\beta,I_{\max}}$.
\end{algorithmic}
\end{algorithm}
\subsection{Proposed Solution for Transmissive RIS}
The two subproblems of transmissive RIS are outlined as follows:
\vspace{-3pt}
\begin{enumerate}
\item For given $\mathbf{S_{\beta\alpha}^{(1)}}$ and $\mathbf{S_{\alpha\beta}^{(2)}}$, solve:
\begin{subequations}
\label{eqn:ao-precod-mat_1_trans}
\begin{align*}
\label{eqn:ao-precod-mata_trans} \tag{44a}
\argmin_{\rho, \f,{\mathbf\Upsilon}} \quad & \norm{\rho\mathbf{H}^\hr_{\rm T}\f -\I_K}?F^2 +K\rho^2\sigma?w^2, \\\
\label{eqn:ao-precod-matb_trans}\tag{44b}
\tn{subject to} \quad & \|\f\|?F^2=P,\\ 
\label{eqn:ao_2_trans} \tag{44c}
& \upsilon_{im}=0, \quad \forall i \neq m,\\
\label{eqn:ao_2_d_trans} \tag{44d}
& |\upsilon_{ii}|=1, \quad \forall i=1,2, \cdots, M. \\
\end{align*}
\end{subequations}
\item For given $\rho$, $\f$, and ${\mathbf\Upsilon}$, solve:
\begin{subequations}
\label{eqn:ao-phase-mat_1_trans}
\begin{align*}
\label{eqn:ao-phase-mata_trans} \tag{45a}
\ds\argmin_{ \mathbf{S_{\alpha\beta}^{(2)}},  \mathbf{S_{\beta\alpha}^{(1)}}} \quad \mathbb{E}_{\h?{r-u},\h?{b-r}} \quad & \Bigl\{\norm{\rho\mathbf{H}^\hr_{\rm T}\f -\I_K}?F^2 +K\rho^2\sigma?w^2\Bigr\}, \\
\label{eqn:eq91-b_trans} \tag{45b}
\tn{subject to} \quad & 
\mathbf{S_{\beta\alpha}^{(1)}}\mathbf{S_{\beta\alpha}^{(1)}}^\hr=\I_{M}, \\
\label{eqn:eq91-c_trans} \tag{45c}
& 
\mathbf{S_{\alpha\beta}^{(2)}}\mathbf{S_{\alpha\beta}^{(2)}}^\hr=\I_{M}.
\end{align*}
\end{subequations}
\end{enumerate}
The first sub-problem is addressed utilizing the method detailed in \cite{9474428} via the VAMP algorithm. Subsequently, the solution to the second sub-problem, which determines the optimal radiation pattern matrices, is obtained by employing the projection gradient descent method, as described below.
\vspace{3pt}
\subsubsection{Projected Gradient Descent Method}
The MSE function for each channel realization corresponding to the transmissive model is presented below:
\small
\begin{equation}
\textit{f}(\mathbf{S_{\alpha\beta}^{(2)}},\mathbf{S_{\beta\alpha}^{(1)}})=  \norm{\rho\mathbf{H}^\hr_{\rm T}\f -\I_K}?F^2 +K\rho^2\sigma?w^2. \tag{46}
\end{equation}
\normalsize
Here also the objective function (\ref{eqn:ao-phase-mata_trans}) contains an expectation involving the channels $\h?{r-u}$ and $\h?{b-r}$. Consequently, Monte Carlo sampling is employed to compute the gradient for each parameter, considering a set of $Q$ channels as samples, similar to the reflective case.
Thereby, the overall gradient of the function with respect to $\mathbf{S_{\beta\alpha}^{(1)}}$ is
\begin{equation}
\label{eq:G_aa_1}
\mathbf{G}_{\mathbf{S}_{\beta\alpha}^{(1)}}=\frac{1}{Q}\sum_{q=1}^Q \left[\pdv{\textit{f}(\mathbf{S_{\alpha\beta}^{(2)}},\mathbf{S_{\beta\alpha}^{(1)}})}{\mathbf{S_{\beta\alpha}^{(1)}}}\right]_\textit{q},\; \mbox{where} \tag{47}
\end{equation}
\normalsize
\begin{align}
\scriptsize
\begin{split}
&\left[\pdv{\textit{f}(\mathbf{S_{\alpha\beta}^{(2)}},\mathbf{S_{\beta\alpha}^{(1)}})}{\mathbf{S_{\beta\alpha}^{(1)}}}\right]_\textit{q}=2\left(\rho_\textit{q}\h?{r-u}^\hr\mathbf{S_{\alpha\beta}^{(2)}}\mathbf{\Upsilon}_\textit{q}\right)^\tr \left(\rho_\textit{q}\h?{r-u}^\hr\mathbf{S_{\alpha\beta}^{(2)}}\mathbf{\Upsilon}_\textit{q}\mathbf{S_{\beta\alpha}^{(1)}}\h?{b-r}\f_\textit{q}-\I_K\right)^* \\
& \left(\h?{b-r}\f_\textit{q}\right)^{\tr},
\end{split}
\tag{48}
\end{align}
\normalsize
in which, similar to the reflection case,  $\mathbf{\Upsilon}_\textit{q}$, $\f_\textit{q}$, $\rho_\textit{q}$ denote optimum phase shift, precoding and scaling parameter values for each channel (sample). Then, the gradient of the function with respect to $\mathbf{S_{\alpha\beta}^{(2)}}$ is given by
\begin{equation}
\label{eq:G_ab_2}
\mathbf{G}_{\mathbf{S}_{\alpha\beta}^{(2)}}=\frac{1}{Q}\sum_{q=1}^Q \left[\pdv{\textit{f}(\mathbf{S_{\alpha\beta}^{(2)}},\mathbf{S_{\beta\alpha}^{(1)}})}{\mathbf{S_{\alpha\beta}^{(2)}}}\right]_\textit{q},  \; \mbox{where}\tag{49}
\end{equation}
\normalsize 
\begin{align}
\scriptsize
\begin{split}
&\left[\pdv{\textit{f}(\mathbf{S_{\alpha\beta}^{(2)}},\mathbf{S_{\beta\alpha}^{(1)}})}{\mathbf{S_{\alpha\beta}^{(2)}}}\right]_\textit{q}=2\left(\rho_\textit{q}\h?{r-u}^\hr\right)^\tr \left(\rho_\textit{q}\h?{r-u}^\hr\mathbf{S_{\alpha\beta}^{(2)}}\mathbf{\Upsilon}_\textit{q}\mathbf{S_{\beta\alpha}^{(1)}}\h?{b-r}\f_\textit{q}-\I_K\right)^* \\
& \left(\mathbf{\Upsilon}_\textit{q}\mathbf{S_{\beta\alpha}^{(1)}}\h?{b-r}\f_\textit{q}\right)^{\tr}.
\end{split}
\tag{50}
\end{align}
\normalsize
Once the gradients are calculated, the next step would be to update the matrices as follows:
$\mathbf{S_{\beta\alpha}^{(1)}}$ and $\mathbf{S_{\alpha\beta}^{(2)}}$.
\begin{align}
\mathbf{\Tilde{S}_{\beta\alpha}^{(1)}}=\mathbf{S_{\beta\alpha}^{(1)}}-\mu\Bigl\{\mathbf{G}^{*}_{\mathbf{S}_{\beta\alpha}^{(1)}}\Bigr\}, \tag{51}
\end{align}
\normalsize
\small
\begin{align}
\mathbf{\Tilde{S}_{\alpha\beta}^{(2)}}=\mathbf{S_{\alpha\beta}^{(2)}}-\mu\Bigl\{\mathbf{G}^{*}_{\mathbf{S}_{\alpha\beta}^{(2)}}\Bigr\}. \tag{52}
\end{align}
\normalsize
Given our operation within the complex domain, we employ Wirtinger derivatives. Consequently, we utilize the conjugate of the gradient as the search direction. Subsequently, we perform the economy size Singular Value Decomposition (SVD) on each updated matrix, as outlined below:
\begin{equation}
    \label{svd:1}
\left[\mathbf{\Tilde{U}_{\beta\alpha}^{(1)}},\mathbf{\Tilde{\Sigma}_{\beta\alpha}^{(1)}},\mathbf{\Tilde{V}_{\beta\alpha}^{(1)}}\right]=SVD\left(\mathbf{\Tilde{S}_{\beta\alpha}^{(1)}}\right), \tag{53}
\end{equation}
\begin{equation}
    \label{svd:2}
\left[\mathbf{\Tilde{U}_{\alpha\beta}^{(2)}},\mathbf{\Tilde{\Sigma}_{\alpha\beta}^{(2)}},\mathbf{\Tilde{V}_{\alpha\beta}^{(2)}}\right]=SVD\left(\mathbf{\Tilde{S}_{\alpha\beta}^{(2)}}\right). \tag{54}
\end{equation}
Next, we proceed with the projection to satisfy the constraints in \eqref{eqn:eq91-b_trans} and \eqref{eqn:eq91-c_trans}.
\small
\begin{align}
\mathbf{S_{\beta\alpha}^{(1)}}=\mathbf{\Tilde{U}_{\beta\alpha}^{(1)}}\left(\mathbf{\Tilde{V}_{\beta\alpha}^{(1)}}\right)^\hr,\tag{55}
\end{align}
\normalsize
\small
\vspace{-12pt}
\begin{align}
\mathbf{S_{\alpha\beta}^{(2)}}=\mathbf{\Tilde{U}_{\alpha\beta}^{(2)}}\left(\mathbf{\Tilde{V}_{\alpha\beta}^{(2)}}\right)^\hr.\tag{56}
\end{align}
\normalsize
The matrices undergo iterative updates until convergence is achieved. The overall algorithm for the transmissive model is given in \textbf{Algorithm} \ref{algo:joint_op_trans}.  

\begin{algorithm}[!t]
\small
\caption{System Optimization for Transmissive RIS}
\label{algo:joint_op_trans}
\vspace{-0.5pt}
\begin{algorithmic}[1]
\renewcommand{\algorithmicrequire}{\textbf{Input:}}
\renewcommand{\algorithmicensure}{\textbf{Output:}}
\REQUIRE $Q$ samples of the channel matrices $\h?{b-r}$ and $\h?{r-u}$
 \vspace{1pt}
\STATE Initialize $\widehat{\mathbf{S}}_{\beta\alpha_{0}}^{(1)}$ and $\widehat{\mathbf{S}}_{\alpha\beta_{0}}^{(2)}$ 
\vspace{1pt}
\FOR{$k=1,2,\ldots,I_{\max}$}
\FOR{$q=1,2,\ldots,Q$}
\STATE Initialize $\mathbf{\Upsilon}_k$.
\STATE Initialize $\rho_k$ and $\f_k$ as described in \cite{9474428}.
\STATE Compute $\mathbf{\Upsilon}_k$ and $\f_k$ solving sub-problem 1) via the method in \cite{9474428}.
\STATE Compute $\left[\pdv{\textit{f}(\mathbf{S}_{\alpha\beta,k}^{(2)},\mathbf{S}_{\beta\alpha,k}^{(1)})}{\mathbf{S}_{\beta\alpha,k}^{(1)}}\right]_\textit{q}$.
\STATE Compute $\left[\pdv{\textit{f}(\mathbf{S}_{\alpha\beta,k}^{(2)},\mathbf{S}_{\beta\alpha,k}^{(1)})}{\mathbf{S}_{\alpha\beta,k}^{(2)}}\right]_{\textit{q}}$.
\ENDFOR
\STATE Compute $\mathbf{G}_{\mathbf{S}_{\beta\alpha,k}^{(1)}}$ using \eqref{eq:G_aa_1}.  
\STATE Compute $\mathbf{G}_{\mathbf{S}_{\alpha\beta,k}^{(2)}}$ using \eqref{eq:G_ab_2}.
\STATE Compute $\mathbf{{S}}_{\beta\alpha,k}^{(1)}={\mathbf{S}}_{\beta\alpha,k-1}^{(1)}-\mu\Bigl\{\mathbf{G}^{*}_{\mathbf{S}_{\beta\alpha,k}^{(1)}}\Bigr\}$
\vspace{2.5pt}
\STATE Compute $\mathbf{{S}}_{\alpha\beta,k}^{(2)}={\mathbf{S}}_{\alpha\beta,k-1}^{(2)}-\mu\Bigl\{\mathbf{G}^{*}_{\mathbf{S}_{\alpha\beta,k}^{(2)}}\Bigr\}$
\vspace{2pt}
\STATE Compute $\left[\mathbf{{U}}_{\beta\alpha,k}^{(1)},\mathbf{{\Sigma}}_{\beta\alpha,k}^{(1)},\mathbf{{V}}_{\beta\alpha,k}^{(1)}\right]=SVD\left(\mathbf{{S}}_{\beta\alpha,k}^{(1)}\right)$
\vspace{2pt}
\STATE Compute 
$\left[\mathbf{{U}}_{\alpha\beta,k}^{(2)},\mathbf{{\Sigma}}_{\alpha\beta,k}^{(2)},\mathbf{{V}}_{\alpha\beta,k}^{(2)}\right]=SVD\left(\mathbf{{S}}_{\alpha\beta,k}^{(2)}\right)$
\vspace{1.5pt}
\STATE Compute $\widehat{\mathbf{S}}_{\beta\alpha,k}^{(1)}=\mathbf{{U}}_{\beta\alpha,k}^{(1)}\left(\mathbf{{V}}_{\beta\alpha,k}^{(1)}\right)^\hr$
\vspace{2pt}
\STATE Compute $\widehat{\mathbf{S}}_{\alpha\beta,k}^{(2)}=\mathbf{{U}}_{\alpha\beta,k}^{(2)}\left(\mathbf{{V}}_{\alpha\beta,k}^{(2)}\right)^\hr$
\ENDFOR
\ENSURE $\rho_{I_{\max}}$, $\f_{I_{\max}}$, $\mathbf{\Upsilon}_{I_{\max}}$, $\mathbf{S}_{\beta\alpha,I_{\max}}^{(1)}$, and $\mathbf{S}_{\alpha\beta,I_{\max}}^{(2)}$.
\normalsize
\vspace{1pt}
\end{algorithmic}
\vspace{1pt}
\end{algorithm}
\subsection{Analysis of Computational Complexity}
\subsubsection{Reflective RIS Setup}
Consider \textbf{Algorithm} \ref{algo:joint_op}. 
Initializing the 2-DFT matrices $\textbf{U}$ and $\textbf{V}$ requires $\mathcal{O}(M^2)$ operations each, summing up to a total of $2\mathcal{O}(M^2)$. The overall computational complexity of the subsequent inner optimization process, aimed at determining the optimal values for $\mathbf{\Upsilon}$, $\rho$, and $\f$ using the method described in \cite{10417011}, entails a total complexity of $Q\bigl(3\mathcal{O}(M^3) + 4\mathcal{O}(N^3) + \mathcal{O}(KNM+KN^2)\bigr)$ \cite{10417011}. Computing the two partial derivatives inside the inner loop incurs a computational complexity of $2Q\mathcal{O}(M^3)$. Subsequently, updating the two matrices $\mathbf{\Sigma_{\alpha\alpha}}$, $\mathbf{\Sigma_{\alpha\beta}}$, and ensuring the symmetry of the matrices involves a total complexity of $4\mathcal{O}(M^2)$. Finally, the two matrix-matrix products related to SVD steps require $2\mathcal{O}(M^3)$ operations.
Hence, the overall complexity is $I_{\max}\Bigl(Q\left(3\mathcal{O}(M^3)+4\mathcal{O}(N^3)+\mathcal{O}(KNM+KN^2)\right)+6\mathcal{O}(M^2)+2\textit(Q+1)\mathcal{O}(M^3)\Bigr)$, where $I_{\max}$ is the number of iterations of outer loop. 
Note that the dominant complexity factor is $\mathcal{O}(M^3)$, primarily due to calculating the matrix inverse. However, despite this seemingly critical complexity, optimization of the scattering parameters on-the-fly helps mitigate its impact.

\subsubsection{Transmissive RIS Setup}
Consider \textbf{Algorithm} \ref{algo:joint_op_trans}.
The overall complexity of computing the optimal $\mathbf{\Upsilon}$, $\rho$, and $\f$, as outlined in \cite{9474428} requires the total complexity of $Q\mathcal{O}(KNM+KN^2)$. Similar to the previous case, computing partial derivatives takes $2Q\mathcal{O}(M^3)$. Then updating the matrices incurs a complexity of $2\mathcal{O}(M^2)$. Finally, calculating the SVD and matrix-matrix multiplication adds a complexity of $4\mathcal{O}(M^3)$. Hence, the overall complexity of the algorithm is $I_{\max}\bigl(Q\mathcal{O}(KNM+KN^2)+2\mathcal{O}(M^2)+2(Q+2)\mathcal{O}(M^3)\bigr)$.
\subsection{Practical Limitations}
In this design, we have assumed a lossless structure and narrow-band systems. To address losses and broaden its applicability, further investigation from an EM perspective is needed. The design itself is still an open area, but it remains a more reasonable option than existing non-local fully-connected complex designs.
\begin{figure*}
\vspace{-10pt}
\centering
\begin{subfigure}{.333\textwidth}
\centering
\includegraphics[scale=0.25]{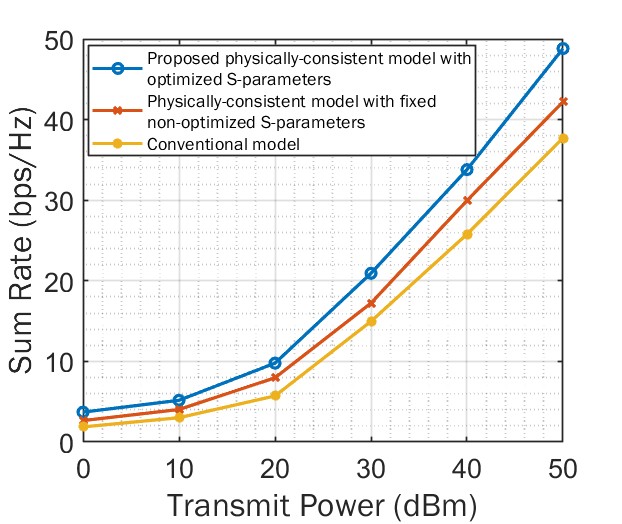}
\caption{\footnotesize $10$ channels, $K=6$, $N=32$ and $M=64$.}
\label{fig:sumrate_64_1}
\end{subfigure}%
\begin{subfigure}{.334\textwidth}
\centering
\includegraphics[scale=0.25]{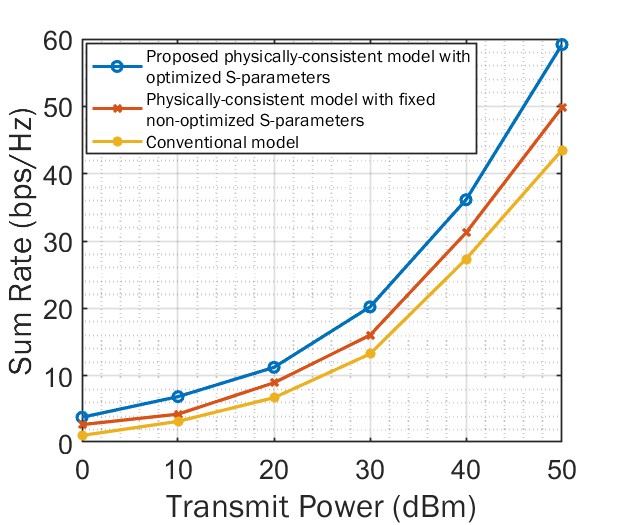}
\caption{\footnotesize $10$ channels, $K=8$, $N=32$ and $M=64$.}
\label{fig:sumrate_64_2}
\end{subfigure}%
\begin{subfigure}{.333\textwidth}
\centering
\includegraphics[scale=0.25]{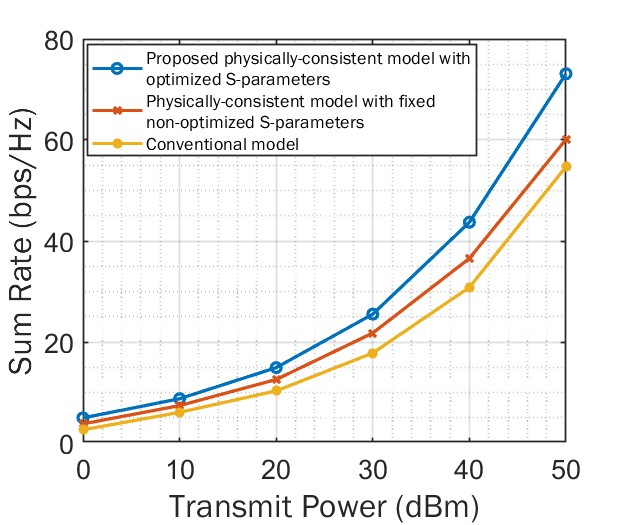}
\caption{\footnotesize $6$ channels, $K=8$, $N=32$ and $M=100$.}
\label{fig:sumrate_64_3}
\end{subfigure}
\caption{Sum-rate performance versus the transmit power $P$ in dBm considering various system parameters using parametric channel modeling for reflective RIS.}
\label{fig:sumrate_64}
\end{figure*}
\begin{figure}[h]
\centering
\includegraphics[scale=0.32]{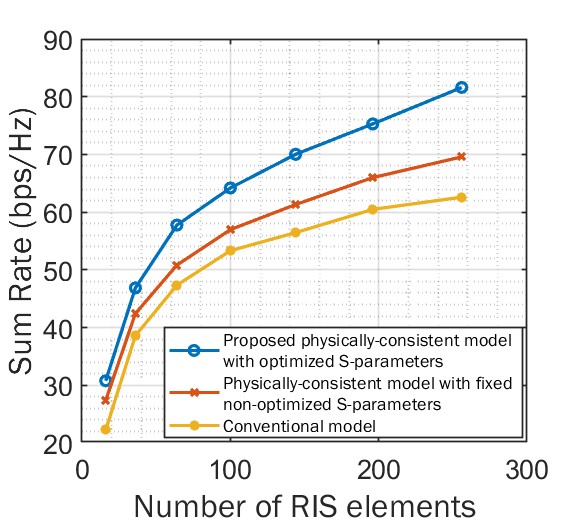}
\caption{Sum-rate performance versus the number of RIS elements $M$ considering $10$ channels with $K=5$, $P=50$ dBm, and $N=32$ using parametric channel modeling for reflective RIS.}
\label{fig:4}
\end{figure}

\section{Numerical Results and Discussion}
We present simulation results for the proposed scheme. The channels $\h?{b-r}$ and $\h?{r-u}$ are characterized using both parametric channel modeling \cite{10096563,10417011} and geometric channel modeling \cite{9732214,9847080} for both RIS models. We present and analyze the simulation results based on these two distinct modeling approaches separately. 

In the setup, a uniform linear array with $N$ antennas is considered for the base station (BS), while a square uniform planar array containing $M$ reflective elements serves for the RIS. The channel between the BS and RIS is characterized by \cite{10096563,10417011}
\begin{equation}
\label{eqn:eq_H_b-r}
\h?{b-r}~=~\sqrt{L(d?{RIS})}\ds\sum_{q=1}^{Q?{b-r}}\cn_q\as?{RIS}(\varphi_q,\psi_q)\as?{BS}(\phi_q)^\tr.
\vspace{2pt} \tag{57}
\end{equation}
The channel vectors corresponding to the links between the RIS and the $k$-th users are as follows \cite{10096563,10417011}:
\small
\begin{equation}
\label{eqn:eq_H_r-u}
\hs_{\tn{r-u},k}~=~\sqrt{L(d'_k)}\ds\sum_{q=1}^{Q?{r-u}}\cn_{k,q}\as?{IRS}(\varphi_{k,q},\psi_{k,q}); \quad k=1, \cdots, K. \tag{58}
\end{equation}
\normalsize
\noindent
Here, the angles $\varphi$ and $\psi$ denote the elevation angle and azimuth angle, respectively. The vectors $\as?{RIS}(\varphi,\psi)$ and $\as?{BS}(\phi)$ represent the array response of the RIS and the BS \cite{10096563,10417011}. The distance-dependent path-loss factor, represented by the term $L(d)$, is defined as follows \cite{10096563,10417011}: 
\small
\begin{equation}
\label{eq:Ddpath_loss}
L(d)=C?0(d/d?0)^{\eta}. \tag{59}
\end{equation}
\normalsize
The sum-rate $C$ is the performance metric \cite{10096563,10417011}. 
Note that, existing non-local RIS structures rely on hardware architectures with higher overhead. Consequently, comparing simulation results between these models and ours may not provide a fair comparison, as both designs are not in the same common ground. Therefore, we consider some other benchmarks for the comparison outlined in the subsequent subsections.  

\subsection{Simulation Results for the Reflective RIS Model}
For benchmarking for the reflective setup, we consider the following two baselines:
\begin{itemize}
    \item The physically-consistent design of \cite{10096563,10417011} that incorporates MC, but considers a fixed non-optimized scattering matrix $\mathbf{S}_{\alpha\alpha}$ and $\mathbf{S}_{\alpha\beta}=\mathbf{I}_M$.
    \item The conventional (hypothetical) model described in \cite{9474428} that does not incorporate MC in the analysis.
\end{itemize}
\vspace{3pt}
\subsubsection{Parametric Channel Modeling Setup}
In this modeling approach, azimuth and elevation angles are chosen randomly. The azimuth angle $\psi_q$ follows a uniform distribution across the interval $[0,2\pi]$, while the elevation angle $\varphi_q$ follows a uniform distribution across the interval $[0,\pi]$. \cite{10096563,10417011}. Table~I illustrates the simulation parameters considered \cite{10096563,10417011}. Note that the distances between the BS and RIS, as well as between the RIS and users, were solely utilized for path loss calculations.

\begin{table}[!t]
\centering
\caption{Simulation parameters}
\label{table:sim-parameter}
{
\begin{tabular} 
{|p{0.25\textwidth}|P{0.13\textwidth}|}
\hline 
\vspace{-10pt} \center{\textbf{Simulation parameter}} & \vspace{-7pt} \textbf{Notation} \\
\hline 

Reference distance & $d?0=1$ m \\
\hline
RIS-BS distance & $d?{RIS}=500$ m\\
\hline
RIS-User distance & $d' \in [10,50]$ m \\
\hline
Channel path numbers of RIS-BS & $Q?{b-r}=8$ \\
\hline
Channel path numbers of RIS-user & $Q?{r-u}=2$ \\
\hline
Channel path gain & $c_q\sim\CG\N(0,1)$ \\
\hline
RIS-BS, RIS-user, path-loss exponent & $\eta=2.5$ \\
\hline
BS-user path-loss exponent & $\eta=3.7$ \\
\hline
Path-loss at the reference distance & $C?0=-30 \ \tn{dB}$ \\
\hline
Noise variance & $\sigma?w^2=-100$ dBm \\
\hline
\end{tabular}
}
\end{table}

It can be observed in Fig. \ref{fig:sumrate_64} that there is an improvement in the sum rate when optimizing MC via the proposed offline approach, as compared to the other two cases. Figure~\ref{fig:4} illustrates the variation of the sum rate versus the $M$ number of RIS elements, and a considerable improvement is observed with the proposed optimization approach. As $M$ increases, the RIS elements need to be placed in close proximity to satisfy a desired surface area. Placing elements closer results in stronger MC. Consequently, an increase in $M$ leads to higher sum-rate performance when optimizing the MC. Therefore, it can be concluded from all simulation results that it suffices for the MC to be optimized offline to gain from its optimization. 
\begin{figure}[h!]
\centering
\includegraphics[scale=0.3]{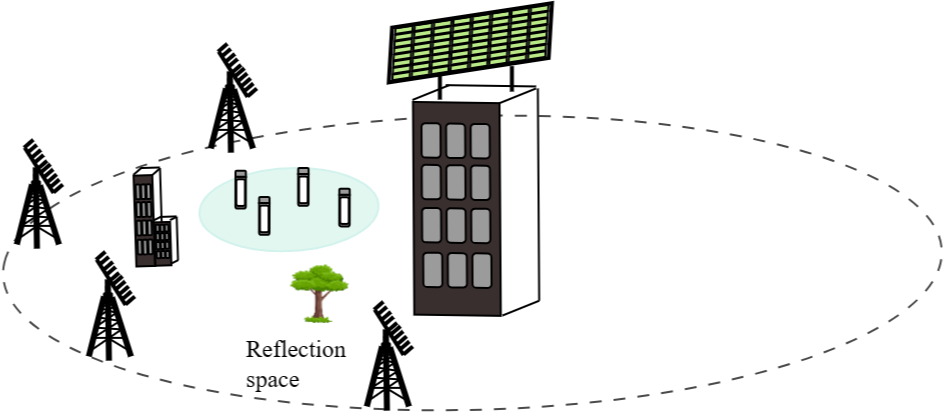}
\caption{Geometric channel modeling setup for reflective RIS  model}
\label{fig:geo}
\end{figure}
\begin{figure*}
\vspace{-10pt}
\centering
\begin{subfigure}{.333\textwidth}
\centering
\includegraphics[scale=0.25]{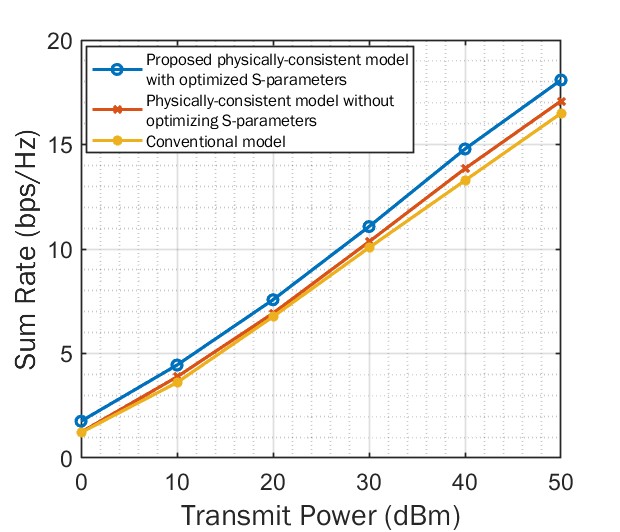}
\caption{\footnotesize $10$ channels, $K=1$, $N=32$ and $M=64$.}
\label{fig:sumrate_64_1_geo}
\end{subfigure}%
\begin{subfigure}{.334\textwidth}
\centering
\includegraphics[scale=0.25]{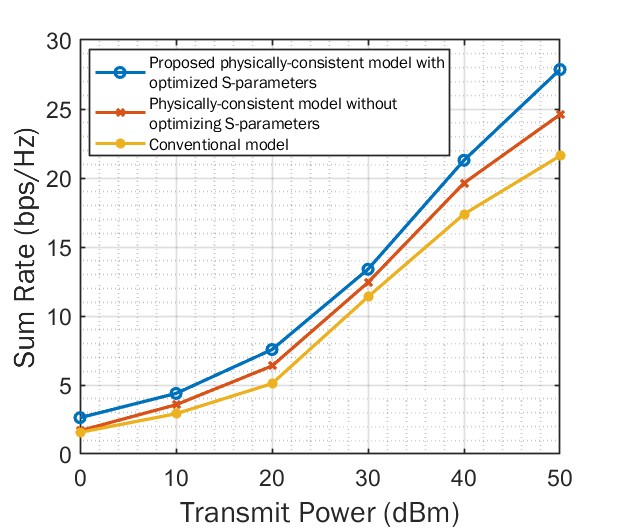}
\caption{\footnotesize $10$ channels, $K=2$, $N=32$ and $M=64$.}
\label{fig:sumrate_64_2_geo}
\end{subfigure}%
\begin{subfigure}{.333\textwidth}
\centering
\includegraphics[scale=0.25]{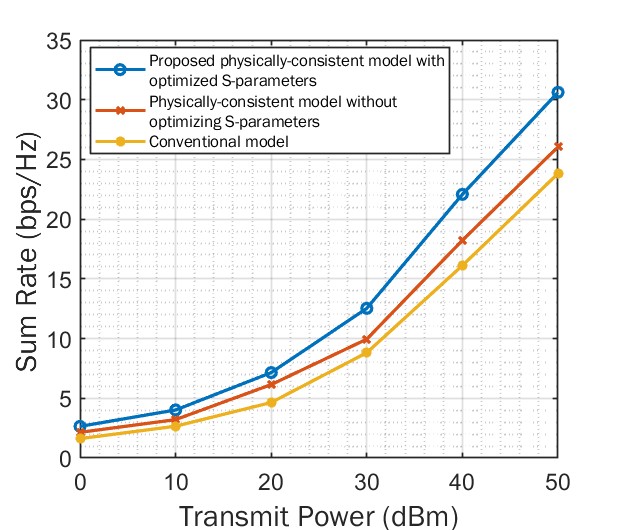}
\caption{\footnotesize $10$ channels, $K=3$, $N=32$ and $M=64$.}
\label{fig:sumrate_64_3_geo}
\end{subfigure}
\caption{Sum-rate performance versus the transmit power $P$ in dBm considering 4 BSs using geometric channel modeling for reflective RIS}
\label{fig:sumrate_64_geo_4BS}
\end{figure*}
\begin{figure*}
\centering
\begin{subfigure}{.333\textwidth}
\centering
\includegraphics[scale=0.25]{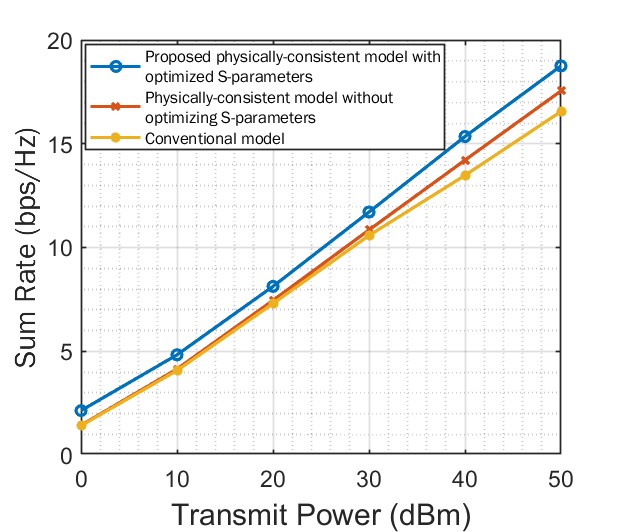}
\caption{\footnotesize $10$ channels, $K=1$, $N=32$ and $M=64$.}
\label{fig:sumrate_64_1_geo_8BS}
\end{subfigure}%
\begin{subfigure}{.334\textwidth}
\centering
\includegraphics[scale=0.25]{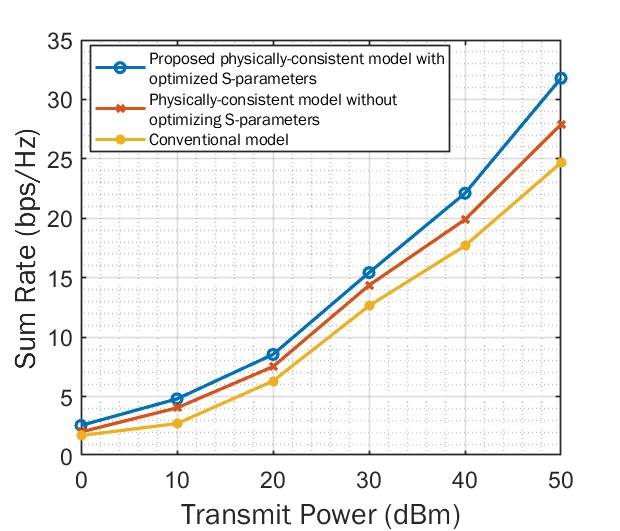}
\caption{\footnotesize $10$ channels, $K=3$, $N=32$ and $M=64$.}
\label{fig:sumrate_64_2_geo_8BS}
\end{subfigure}%
\begin{subfigure}{.333\textwidth}
\centering
\includegraphics[scale=0.25]{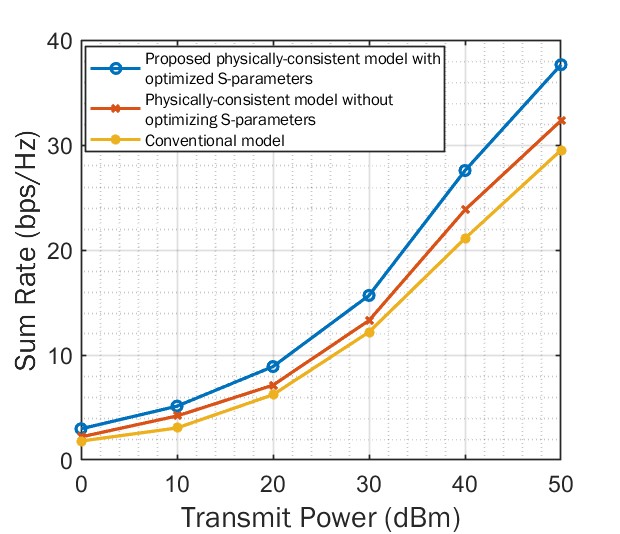}
\caption{\footnotesize $10$ channels, $K=4$, $N=32$ and $M=64$.}
\label{fig:sumrate_64_3_geo_8BS}
\end{subfigure}
\caption{Sum-rate performance versus the transmit power $P$ in dBm considering 8 BSs using geometric channel modeling for reflective RIS.}
\label{fig:sumrate_64_geo_8BS}
\end{figure*}
\vspace{3pt}
\subsubsection{Geometric Channel Modeling Setup}
In this modeling framework, azimuth and elevation angles are computed based on the geometric positions of the BS, RIS, and users, ensuring the angles align with the geometry of the simulation setup. 
Instead of using a predetermined number of channel paths for the BS-RIS, we introduce multiple BSs. This transforms the setup into a distributed multiple-input multiple-output (MIMO) scenario \cite{10323184,6601775}, offering increased flexibility and adaptability. In this scenario, BSs are positioned along the circumference of a circle with a radius of 500m from the RIS and users are situated within an area ranging from 10m to 50m distance from the RIS, as shown in Fig. \ref{fig:geo}. Figures \ref{fig:sumrate_64_geo_4BS} and \ref{fig:sumrate_64_geo_8BS} show the variation of sum-rate with power for 4 BSs and 8 BSs respectively. As the number of BSs in the setup increases, the capacity to serve more users concurrently improves. Furthermore, we notice an improvement in performance with the proposed model, even when utilizing the geometric model.
\subsection{Simulation Results for the Transmissive RIS Model}
Since we assume no MC in the transmissive model, we use the following benchmarking to compare the results. 
\begin{itemize}
    \item The conventional (hypothetical) model described in \cite{9474428} that does not incorporate MC in the analysis.
\end{itemize}
\vspace{3pt}
\subsubsection{Parametric Channel Modeling}
\begin{figure*}
\vspace{-10pt}
\centering
\begin{subfigure}{.333\textwidth}
\centering
\includegraphics[scale=0.25]{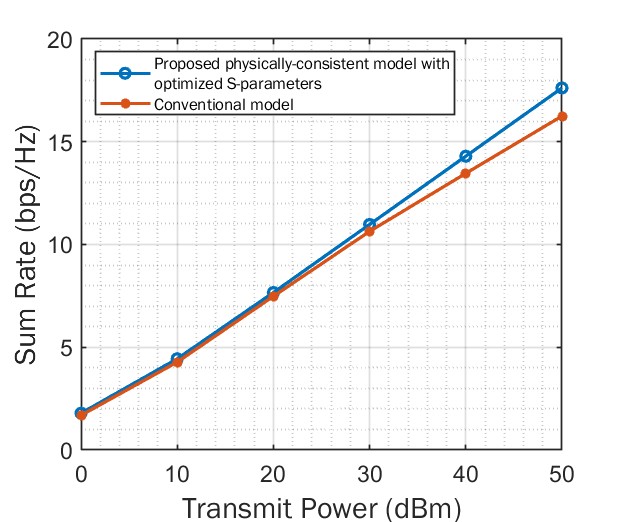}
\caption{\footnotesize $10$ channels, $K=1$, $N=32$ and $M=64$.}
\label{fig:sumrate_64_1_para_trans}
\end{subfigure}%
\begin{subfigure}{.334\textwidth}
\centering
\includegraphics[scale=0.25]{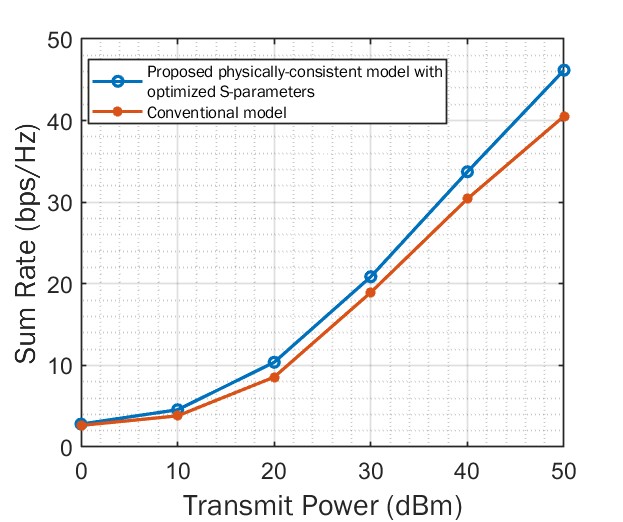}
\caption{\footnotesize $10$ channels, $K=4$, $N=32$ and $M=64$.}
\label{fig:sumrate_64_4_para_trans}
\end{subfigure}%
\begin{subfigure}{.333\textwidth}
\centering
\includegraphics[scale=0.25]{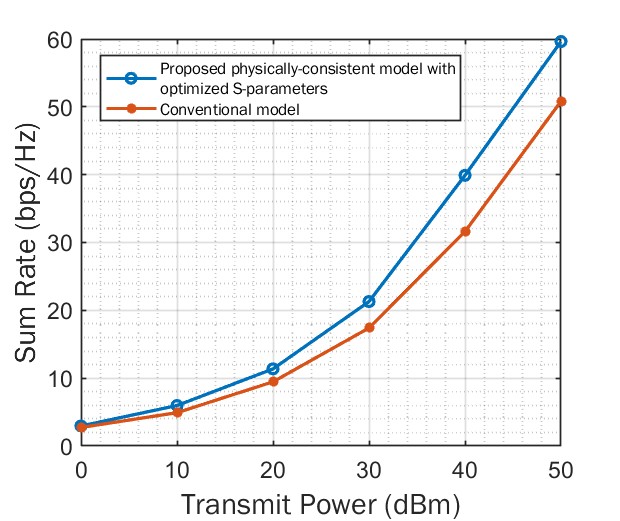}
\caption{\footnotesize $10$ channels, $K=8$, $N=32$ and $M=64$.}
\label{fig:sumrate_64_8_para_trans}
\end{subfigure}
\caption{Sum-rate performance versus the transmit power $P$ in dBm considering various parameters using parametric channel modeling for transmissive RIS.}
\label{fig:sumrate_64_para_trans}
\end{figure*}

\begin{figure}[h]
\centering
\includegraphics[scale=0.31]{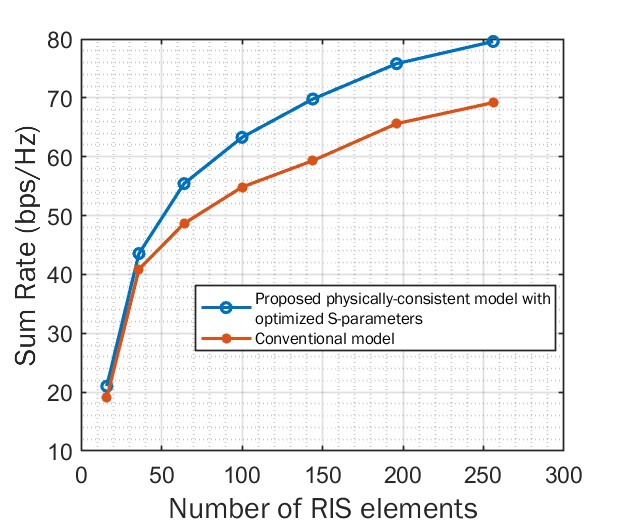}
\caption{Sum-rate performance versus the number of RIS elements $M$ considering $10$ channels with $K=6$, $P=50$ dBm, and $N=32$ using parametric channel modeling for transmissive RIS.}
\label{fig:4_trans_RIS}
\end{figure}

\begin{figure}[h]
\centering
\includegraphics[scale=0.31]{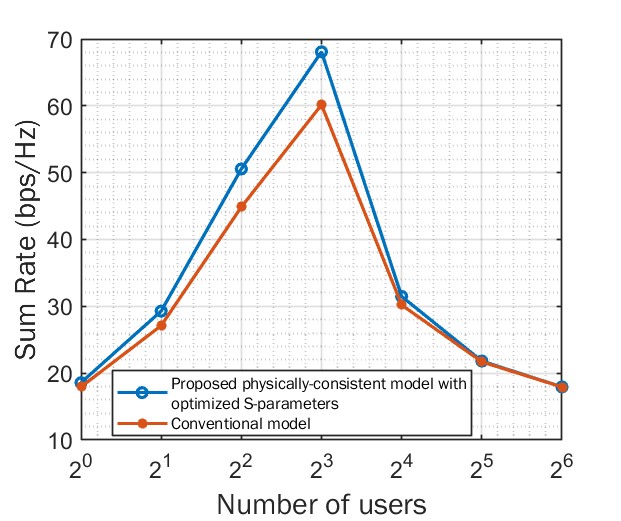}
\caption{Sum-rate performance versus the number users considering $10$ channels with $M=100$, $P=50$ dBm, and $N=32$ using parametric channel modeling for transmissive RIS.}
\label{fig:4_trans_users}
\end{figure}

Here, we employ the same simulation parameters as outlined in Table \ref{table:sim-parameter}. Figure \ref{fig:sumrate_64_para_trans} shows the variation of sum-rate with transmit power for the transmissive model. We can observe a significant amount of gain in the proposed model compared to the conventional model. Furthermore, a noticeable improvement in the sum-rate is evident when adjusting the number of RIS elements, as illustrated in the Fig. \ref{fig:4_trans_RIS}. Figure \ref{fig:4_trans_users} illustrates the sum-rate variation with the number of users. Notably, the gain rises with the user count up to 8, and then declines. This behavior is attributed to the utilization of 8 RIS-BS multi-paths. Consequently, the system cannot support more than 8 users. Having additional RIS-BS multi-paths would enable us to support for a larger number of users. These results indicate that we can achieve an improvement compared to the conventional model, even without incorporating mutual coupling. Additionally, solely optimizing radiation patterns would be sufficient to attain an optimized non-local RIS.
\vspace{3pt}
\subsubsection{Geometric Channel Modeling} 
\begin{figure*}[ht!]
\vspace{-10pt}
\centering
\begin{subfigure}{.333\textwidth}
\centering
\includegraphics[scale=0.25]{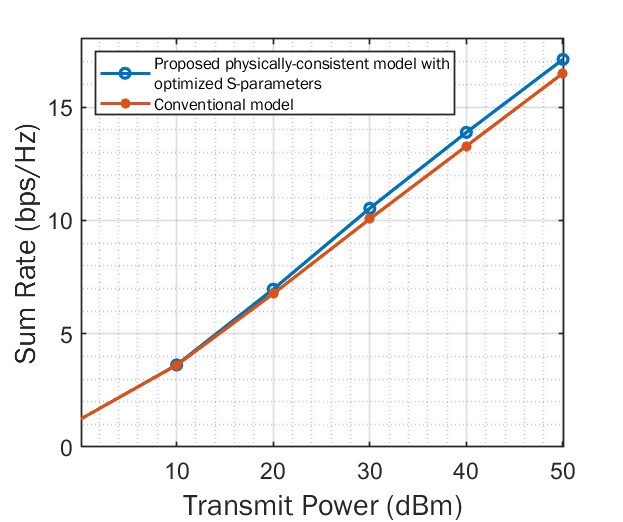}
\caption{\footnotesize $10$ channels, $K=1$, $N=32$ and $M=64$.}
\label{fig:sumrate_64_1_geo_trans}
\end{subfigure}%
\begin{subfigure}{.334\textwidth}
\centering
\includegraphics[scale=0.25]{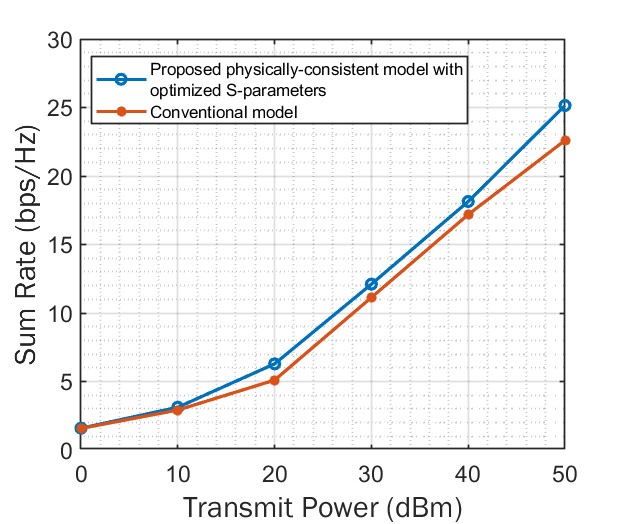}
\caption{\footnotesize $10$ channels, $K=2$, $N=32$ and $M=64$.}
\label{fig:sumrate_64_2_geo_trans}
\end{subfigure}%
\begin{subfigure}{.333\textwidth}
\centering
\includegraphics[scale=0.25]{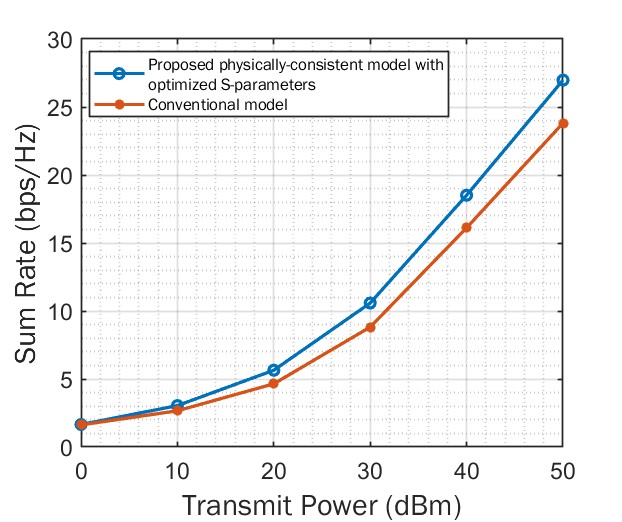}
\caption{\footnotesize $10$ channels, $K=3$, $N=32$ and $M=64$.}
\label{fig:sumrate_64_3_geo_trans}
\end{subfigure}
\caption{Sum-rate performance versus the transmit power $P$ in dBm considering 4 BSs using geometric channel modeling for transmissive RIS.}
\label{fig:sumrate_64_geo_trans}
\end{figure*}

Utilizing the identical geometric setup as in the reflective mode, but relocating users to the transmissive space. The results showcased in Fig. \ref{fig:sumrate_64_geo_trans} demonstrate a considerable improvement in the sum-rate with our proposed scheme compared to the hypothetical model with the geometric channel modeling. 

\section{Conclusion}
We have proposed a novel non-local RIS framework by optimizing the RIS phase shifters and BS active precoders using a physically-consistent model incorporating MC and radiation patterns. We have provided distinct solution approaches considering separately a reflective and a transmissive RIS models. The MC and radiation patterns were optimized through an offline optimization method applied to a class of channels, as opposed to state-of-the-art approaches where no MC and radiation pattern optimization are performed. The resultant nested optimization problem has been decomposed into two sub-problems. The RIS phase shifters and BS active precoding have been designed using online optimization. Our approach emphasizes the potential of ``engineering'' MC and radiation patterns to enhance sum-rate performance without requiring on-the-fly adjustments. Simulation results have been presented considering  both a parametric and a geometric channel model, which validated the effectiveness of the proposed scheme, demonstrating enhanced system performance through active and passive beamforming, incorporating optimized MC and radiation patterns. Furthermore, it was verified that is feasible to achieve non-local RIS configurations, incorporating optimized MC and radiation patterns, all without requiring increased overhead.

\bibliographystyle{IEEEtran}

\vspace{-1mm}
\bibliography{bib}

\end{document}